\documentclass[pra, 10pt, letterpaper,oneside,notitlepage, twocolumn]{revtex4-1}
\usepackage[latin1]{inputenc}
\usepackage{mathtools}
\usepackage{amsfonts}
\usepackage{array}
\usepackage[T1]{fontenc}
\usepackage{amssymb}
\usepackage[left=.75in, right=.75in, top=1in, bottom=1in]{geometry}
\usepackage{amssymb}
\usepackage{graphics}
\usepackage{graphicx}
\usepackage{array}
\usepackage{color}
\usepackage[hypertex, colorlinks, colorlinks=false, linkcolor=blue, citecolor=green, filecolor=black, urlcolor=cyan]{hyperref}

\begin{document}

\title{Physics of the fundamental limits of nonlinear optics:  A theoretical perspective}
\author{Rick Lytel}
\affiliation{Department of Physics and Astronomy, Washington State University, Pullman, Washington  99164-2814}

\begin{abstract}
The theory of the fundamental limits (TFL) of nonlinear optics is a powerful tool for experimentalists seeking to create molecules and materials with large responses, and for theorists who are seeking to understand how the basic elements of quantum theory delineate the boundaries within which these searches should be conducted.  On a practical level, the TFL provides a metric for measuring the performance or 'goodness' of new molecules, relative to what is possible.  Explorations of large sets of structures within the theory provide insight into new design rules for creating more active molecules.  This article is a review of the TFL, starting with a history of its development and its first use to discover that all molecules as of the year 2000 fell a factor of 30 below the limits, and continuing to the present day where the theory continues to provide research opportunities and challenges.  The review focuses on off-resonant nonlinear optics in order to sharply focus on the key elements of the TFL, but pointers are provided to the literature for near- and on-resonance applications.
\end{abstract}



\maketitle

\section{Introduction}\label{sec:intro}
Nonlinear optics is the study of quantum systems with polarizations that are nonlinear functions of external electromagnetic fields.  It is about a year younger than the age of the laser, a little over fifty years old. Lasers are ubiquitous in daily life.  Nonlinear optics has offered scientific research numerous methods for studying materials and measuring their quantum properties, but nonlinear optical materials with large optical responses and low loss, required for most devices, remain elusive\cite{garmi13.01}.  The scientific field has spanned decades of fundamental and applied research on the interactions of strong electromagnetic fields with naturally occurring solid\cite{baugh78.06,lipsc81.01,hugga87.01,cheml80.01}, liquid\cite{giord65.01,giord67.01,ho79.01,chen88.01}, liquid crystal\cite{barni83.01,chen88.01} or gaseous materials\cite{shelt82.01,kaatz98.01}, as well as photonic crystals, mesoscopic solid state wires\cite{guble99.01,foste08.01,tian09.01}, and other artificial systems\cite{cheml85.01,miller1984band,rink89.01,schmi87.01}.  The theory of nonlinear optics is well-established and has recently been reviewed in the literature\cite{kuzyk13.02,kuzyk14.02}.  This paper is a concise review of our present understanding of the physics of the theory of the fundamental limits of nonlinear optics with an eye toward understanding why it is that nearly all molecular systems fall short of the quantum limits and how this discovery, made in 2000 by M.G. Kuzyk\cite{kuzyk00.01} has advanced our understanding of the origin of the limits themselves and the design rules required for the molecular response to approach the limits.

The nonlinear optical response of a material is generated by the collective response of the basic elements comprising it.  Semiconductor nonlinear optics results from excitation of electrons in compound materials\cite{mille81.01}, while the response of a dye doped polymer results from the collective response of the moieties embedded into the polymer matrix\cite{kuzyk94.01,kuzyk06.06}. This review concerns the maximum values of the nonlinear optical response of a molecular-scale structure, not a material.  The ground state dipole moment in the presence of an electric field $\mathcal{E}_{i}$ is expressed as
\begin{equation}\label{polVec}
p_{i}=\mu_{i}+\alpha_{ij}\mathcal{E}_{j}+\beta_{ijk}\mathcal{E}_{j}\mathcal{E}_{k}+\gamma_{ijkl}\mathcal{E}_{j}\mathcal{E}_{k}\mathcal{E}_{l}+\ldots
\end{equation}
where $\mu_{i}$ is ith component of the unperturbed ground state dipole moment vector, $\alpha_{ij}$ is the linear polarizability tensor, $\beta_{ijk}$ is the first hyperpolarizability tensor\cite{kuzyk00.01}, and $\gamma_{ijkl}$ is the second hyperpolarizability tensor\cite{kuzyk00.02}.  In Eqn \ref{polVec} and throughout this paper, repeated tensor indices are summed.  The tensors contain all of the physics of the interaction of a molecule's nuclei and electrons with light or a static electric field, and their tensor structure reflects the rotational properties of the molecule under the proper orthogonal rotation group $O(3)$, as well as inversion through some origin\cite{orr71.01,lytel13.01}. The tensors are calculable using perturbation theory, finite fields, and other numerical methods described later in this review.

In perturbation theory, the three tensors depend on the energy levels, the dipole transition matrix elements, and the linewidths of the levels.  In this review, we denote the energies by $E_{n}$ and most often use the energy difference $E_{n0}=E_{n}-E_{0}$ between the excited state n and the ground state.  The transition moments are expressed as $r^{i}_{nm}=\langle n|r^{i}|m\rangle$, where $i=x,y,z$.  Finally, the natural linewidths are generally a complex function of the quantum properties of the states, but may be written using Fermi's Golden Rule\cite{schif68.01} as a product of the cube of the energy and the absolute square of the transition moment.  In this review, we'll focus on off-resonance responses, as it clears away some of the complexity that can obscure the essential physics of limit theory without loss of generality.  References to the general case are provided as we go along.

The computation of the linear polarizability, and the first and second hyperpolarizabilities, in perturbation theory, was first published by Orr and Ward\cite{orr71.01}, as a sum over unperturbed states.  Symmetries of the structures provide constraints on the nonlinear response and can enhance, reduce, or zero it. Topological properties generally affect the nature of the energy spectrum and its dependence on the eigenmode number.  Geometrical properties primarily affect the magnitude of the transition moments and the nature of the sum over states. The scientist synthesizing new molecules with the goal of achieving large responses at specific optical frequencies has an array of options with which to work.  An additional key element is the scale of the molecule:  Larger, self-similar structures might have larger responses, but materials using these elements have lower number densities of them, resulting in limited benefit to the use of size as a design element.  For this reason, it is imperative to examine the large design space for molecular structures in a scale-independent way.  This immediately raises two questions:  How large can this \emph{intrinsic} response be, and how well do present molecular elements perform?  The theory of fundamental limits is a body of work created specifically to answer these questions\cite{kuzyk00.01,kuzyk00.02,kuzyk03.02}.

A basic question the reader should be asking is why there are any limits at all.  To answer this in a simple way, consider the linear polarizability $\alpha_{ij}$ whose sum over states in the off-resonant limit is given by
\begin{equation}\label{alphaij}
\alpha_{ij} = e^2P_{ij} {\sum_{n}}' \frac{r_{0n}^i r_{n0}^j}{E_{n0}},
\end{equation}
where $P_{ij}$ is a permutation operator over the two cartesian indices and the prime on the sum means skip $n=0$.  This tensor is a sum of terms containing transition moments and energy levels.  The spectra and eigenstates are the eigenvalues and eigenvectors found by diagonalizing the Hamiltonian $H(r^{i},p^{j})$, and the transition moments $r_{nm}^{i}\equiv\langle n|r^{i}|m\rangle$ are computed from the eigenstates.  Spectra and moments are intimately related and not independent of one another.  This relation is expressed concisely through the Thomas-Reiche-Kuhn (TRK) sum rules\cite{thom25.01,reich25.01,kuhn25.01}.  The TRK sum rules are a consequence of the commutators between the position operators and the commutator of those operators with the Hamiltonian $H(r^{i},p^{j})$, where the position and momentum operators satisfy the canonical commutation relations $[r^{i},p^{j}]=i\hbar\delta_{ij}$. The Hamiltonian may depend on other parameters, such as external electromagnetic potentials, but if it satisfies the basic commutator $[r^{i},[r^{j},H]]=-(\hbar^2/m)\delta_{ij}$ \emph{and} its eigenstates satisfy completeness and closure, then by inserting a complete set of states between the operators in the double commutator, one gets the TRK sum rules.  For the x component, these are
\begin{equation}\label{TRKsumrule}
S_{nm} = \sum_{p=0}^{\infty} E_{nm}^{p} x_{np}x_{pm}=\frac{\hbar^2 N}{m}\delta_{nm},
\end{equation}
where
\begin{equation}\label{Epnm}
E_{nm}^{p}=\left[2E_{p0}-(E_{n0}+E_{m0})\right]=E_{pn}+E_{pm}
\end{equation}
The summation spans the complete set of eigenstates of the system, including both discrete and continuum \cite{Bethe77.01} states, and degenerate and non-degenerate states \cite{ferna02.01}. In Eqn \ref{TRKsumrule}, $N$ is the number of participating electrons, and $m$ is the electron mass.

The commutator $[r^{i},H]=i\hbar p^{i}/m$ is familiar to students of introductory quantum theory, where $H(r,p)$ is often written as $p^2/2m+V(r)$ in the presence of a potential function.  But the sum rule commutator $[r^{i},[r^{j},H]]=-(\hbar^2/m)\delta_{ij}$ holds for a larger class of Hamiltonians that include interactions with electromagnetic fields, and a rather novel set of so-called \emph{exotic} Hamiltonians\cite{watki12.01}.  This would seem to encompass all possible physical non-relativistic Hamiltonians.

To see the impact of the sum rules on Eqn \ref{alphaij}, consider the xx component.  Eqn \ref{alphaij} becomes
\begin{eqnarray}\label{alphaxx}
\alpha_{xx} &=& 2e^2{\sum_{n}}' \frac{|x_{0n}|^2}{E_{n0}} \\
&=& 2e^2{\sum_{n}}' \frac{E_{n0}|x_{0n}|^2}{E_{n0}^2}\nonumber \\
&\leq& \frac{2e^2}{E_{10}^2}{\sum_{n}}' E_{n0}|x_{0n}|^2\nonumber
\end{eqnarray}
But from Eqn \ref{Epnm} with $n=m=0$, we have $E_{00}^{p}=2E_{p0}$, and Eqn \ref{TRKsumrule} yields
\begin{equation}\label{TRK00}
S_{00} = 2\sum_{p=0}^{\infty}E_{p0}|x_{p0}|^2=\frac{\hbar^2 N}{m}.
\end{equation}
Using Eqn \ref{TRK00} in Eqn \ref{alphaxx} gives us an upper limit on the x-diagonal polarizability tensor:
\begin{equation}\label{alphaxxL}
\alpha_{xx} \leq \alpha_{max} = \frac{e^2\hbar^2}{m}\frac{N}{E_{10}^2}.
\end{equation}
Thus, the sum rules have acted as constraints to reveal an upper limit to polarizability tensor that depends solely on the energy level splitting between the ground and first excited state, and the number of participating electrons.  This is an exact result.  The sum rules are an infinite set of equations in an infinite number of variables.  In principle, one could solve $H|n\rangle = E_{n}|n\rangle$ for the stationary states of the structure, compute the transition moments, and verify the validity of Eqn \ref{TRKsumrule}.  The veracity of the sum rules is inviolate, making them an exceptional tool for verifying computations.

But they are evidently a tool for constraining the spectra and moments appearing in sum over states expressions, as just demonstrated for the first polarizability tensor. Kuzyk pioneered the use of sum rules for the purpose of examining their effect as constraints on the nonlinear optical tensors, leading for the first time to a quantitative statement of the limits of the size of nonlinear optical tensors, which revealed that all experimental work known up to the time of publication of this work fell a factor of 30 below the limits\cite{kuzyk00.01,kuzyk03.02}.  This discovery, its history, progress in its development, and open questions are the subject of this review.

Section \ref{sec:history} details the development of the theory of fundamental limits (TFL).  The discussion mirrors the original method to elucidate subtleties that are often overlooked by those using the TFL.  The three-level approximation for $\beta$ is introduced and its reduction to the three-level model via truncated sum rules is described.  The theory revealed a gap of a factor of 30 between all experiments as of the year 2000 and the limits, the so-called Kuzyk gap.  The validity of the TFL is discussed, and the concept of the three-level Ansatz (TLA) is introduced and its role in the theory is described.  Scaling and universality in the TFL is summarized and its meaning explored.  The section closes with a history of experiments and models targeted at understanding the limits and to close the Kuzyk gap in the post-TFL years.

Section \ref{sec:results} presents theoretical advances toward understanding the limits and achieving them in molecular systems.  Theoretical advances include potential optimization and Monte Carlo simulations, which result in identifying key features of the spectrum of a good molecule.  The two methods predict different upper limits for the intrinsic nonlinearities, creating a new sort of gap and new challenges to understand it. Most models studied are one-dimensional.  The application of the TFL to quasi-one dimensional quantum graphs that exist in two spatial dimensions is invoked to illustrate that the three-level Ansatz holds for $\beta$ but does not hold for $\gamma$, which requires four states, despite the fact that the maximum value of $\gamma$ is derived using only three states.

Section \ref{sec:issues} addresses outstanding issues in the theory of fundamental limits.  The first concerns the gap in predicted limits between physical systems described by a generalized Hamiltonian (and its spectra and states) with the canonical commutation relation required to generate sum rules, and the limits predicted by using the sum rules alone and selecting spectra and moments at random with sum rules as constraints.  The second concerns the TLA and its relationship to the TFL.  A third concerns new design methods gleaned from the TFL.  A fourth is the application of the TFL to device figures of merit.  The section closes with a concise analysis of what we know about the TFL today, and what remains to be discovered.

Section \ref{sec:end} closes the review with a summary of the main results and a personal perspective based upon several years of work on the theory of fundamental limits using quantum graphs.

The author emphasizes that this article is a brief review of the theory of the fundamental limits of nonlinear optics from his own perspective, for the purposes of providing a framework for readers of the papers in this special issue.  A comprehensive review may be found in reference \cite{kuzyk13.01}.

\section{A brief history of significant research}\label{sec:history}
The theory of the fundamental limits of nonlinear optics was first developed by Kuzyk over fifteen years ago\cite{kuzyk00.01} for the first hyperpolarizability tensor and later extended by him to the second hyperpolarizability tensor\cite{kuzyk00.02} for off-resonant nonlinear optics. The approach was not without controversy\cite{champ05.01,kuzyk05.01}, as it used truncated sum rules to compute an upper limit to $\beta$, and lower and upper limits to $\gamma$.  Fifteen years after his breakthrough efforts, this topic was thoroughly explored by Kuzyk in a tour du force work on handling pathologies in truncated sum rules\cite{kuzyk14.01}. But fundamental issues remain and will be discussed in Section \ref{sec:issues}.

\subsection{The fundamental limits}\label{sec:3level}
An exact computation of the limits imposed by the TFL starts with the full tensor expressions of the hyperpolarizabilities, written in first order perturbation theory as a sum over states\cite{orr71.01}.  Far from resonance, the tensor components of the first hyperpolarizability, $\beta_{ijk}$ are given by
\begin{equation}\label{betaijk}
\beta_{ijk} = \frac{e^3}{2} P_{ijk} {\sum_{n,m}}' \frac{r_{0n}^i \bar{r}_{nm}^j r_{m0}^k}{E_{n0} E_{m0}}
\end{equation}
where the prime indicates that the ground state is excluded from the summation, the superscripts $i$, $j$ and $k$ can take on $x$, $y$ or $z$ (the Cartesian components), $P_{ijk}$ permutes all the indices in the expression, $\bar{r}_{nm} = r_{nm} - r_{00}\delta_{nm}$, and $E_{nm} = E_n - E_m$ is the energy between two eigenstates $n$ and $m$.  Given the similarity of this form to that for $\alpha_{ij}$ in Eqn \ref{alphaij}, it seems reasonable to expect that sum rules of the form Eqn \ref{TRKsumrule} may provide sufficient constraints to restrict the maximum value of any of the tensor components in Eqn \ref{betaijk}.  This was first demonstrated by Kuzyk\cite{kuzyk00.01}, in a method detailed here.

Consider the x-diagonal component of $\beta_{ijk}$ and call it $\beta$ for simplicity.  Then we may write
\begin{eqnarray}\label{beta3L}
\beta &=& 3e^3 \left(\frac{|x_{01}|^{2}\bar{x}_{11}}{E_{10}^{2}}+\frac{|x_{02}|^{2}\bar{x}_{22}}{E_{20}^{2}}+\left[\frac{x_{01}x_{20}x_{12}}{E_{10}E_{20}}+c.c.\right]+\cdots\right)\nonumber \\ &\equiv& \beta_{11} + \beta_{22} + [\beta_{12}+c.c]+\cdots\equiv \beta_{3L}+\cdots
\end{eqnarray}
which defines $\beta_{3L}$ as the sum over states arising from transitions involving only three energy levels, and $\beta=\beta_{3L}$ plus contributions from all terms with higher levels.  The first two terms are dipolar terms, expressible as a $J=1$ spherical tensor\cite{lytel13.01}, where as the third term is an octupolar term with $J=3$.  A priori, there is no reason to assume that three terms would be sufficient to compute $\beta$ under any circumstances--it is simply a partial sum of $\beta$ that involves only the first three levels.  The energy differences and transition moments in $\beta$ and $\beta_{3L}$ satisfy the full TRK sum rules.

To derive a limit, Kuzyk\cite{kuzyk00.01} truncated the sum rules to three-levels, too, and then wrote each of the three terms in $\beta_{3L}$ as functions of $x_{01}$ alone as follows. From Eqn \ref{TRKsumrule}, we find
\begin{equation}\label{S00}
S_{00}=\hbar^2 N/m\approx 2[|x_{01}|^2E_{10}+|x_{02}|^2E_{20}],
\end{equation}
\begin{equation}\label{S11}
S_{11}=\hbar^2 N/m\approx 2[-|x_{01}|^2E_{10}+|x_{12}|^2E_{21}],
\end{equation}
\begin{equation}\label{S01}
S_{01}=0\approx x_{01}\bar{x}_{11}E_{10}+x_{02}x_{21}(E_{21}+E_{20}),
\end{equation}
and
\begin{equation}\label{S02}
S_{02}=0\approx  x_{02}\bar{x}_{22}E_{20}+x_{01}x_{12}(E_{10}-E_{21}),
\end{equation}
where $\bar{x}_{nn}\equiv x_{nn}-x_{00}$ is the shift in the dipole moment.  All of the equalities in Eqns \ref{S00}-\ref{S02} are approximate, and there is no a priori reason to assume that the remaining terms on the right-hand side of these equations are vanishingly small.  However, the sum rules do converge, so eventually the general term of each asymptotes to zero.

[As an aside, we note that the original Letter computed the upper limit to $\beta$ in a two state model by taking the limit of the approach we are following as $E_{20}\rightarrow\infty$.  This limit exists, but the two level model, with sum rules truncated to two states, requires either $x_{01}=0$ or $\bar{x}_{10}=0$ and predicts $\beta$ is exactly zero.  We are obligated, then, to consider at least three states when seeking an upper bound for $\beta$.]

We can use the exact form of the sum rule $S_{00}$ to write
\begin{equation}\label{x01MAX}
|x_{01}|^2=\frac{\hbar^2 N}{2mE_{10}}-\sum_{n=2}^{\infty}\frac{E_{n0}}{E_{10}}|x_{n0}|^2<|x_{max}|^2
\end{equation}
which defines a maximum value of $x_{01}$ as
\begin{equation}
x_{max}=\sqrt{\frac{\hbar^2 N}{2mE_{10}}}.
\end{equation}
With this definition, Eqns \ref{S00}-\ref{S02} may be used to write each term in the three-level expression, Eqn \ref{beta3L}, for $\beta$ in terms of $x_{01}$ and its maximum value, $x_{max}$, and the ratio $E\equiv E_{10}/E_{20}$.  We find that
\begin{equation}\label{x02}
|x_{02}|=\sqrt{E\left(|x_{max}|^2-|x_{01}|^2\right)},
\end{equation}
\begin{equation}\label{x12}
|x_{12}|=\sqrt{\frac{E}{1-E}\left(|x_{max}|^2+|x_{01}|^2\right)},
\end{equation}
\begin{equation}\label{x01xbar11}
|x_{01}|\bar{x}_{11}=\frac{2-E}{\sqrt{1-E}}\sqrt{\left(|x_{max}|^4-|x_{01}|^4\right)},
\end{equation}
and
\begin{equation}\label{x02xbar22}
|x_{02}|\bar{x}_{22}=(2E-1)|x_{01}|\sqrt{\frac{E}{1-E}\left(|x_{max}|^2+|x_{01}|^2\right)}.
\end{equation}
Finally, we may multiply Eqn \ref{S02} by $x_{20}$ to get $x_{01}x_{12}x_{20}=|x_{20}|^2\bar{x}_{22}/(1-2E)$, which may be written using Eqns \ref{x02} and \ref{x02xbar22} as
\begin{equation}\label{x01x12x20}
x_{01}x_{12}x_{20}=-|x_{01}|\sqrt{\frac{E^2}{1-E}\left(|x_{max}|^4-|x_{01}|^4\right)}.
\end{equation}
Substituting Eqns \ref{x02}-\ref{x01x12x20} into Eqn \ref{beta3L} yields the \emph{three-level model prediction $\beta_{3L}(ext)$ for $\beta$ when the SOS and the sum rules are truncated to three levels}:
\begin{eqnarray}\label{beta3Lext}
\beta_{3L}&\rightarrow& \beta_{3L}(ext)=\beta_{max}f(E)G(X),\nonumber \\
f(E)&=&\left[(1-E)^{3/2} \left( E^2 + \frac {3} {2} E + 1 \right)\right], \\
G(X)&=&\left[\sqrt[4]{3} X \sqrt{\frac {3} {2} \left( 1 - X^4\right)}\right]\nonumber
\end{eqnarray}
with
\begin{equation}\label{betaMax}
\beta_{max}=3^{1/4}\left(\frac{e\hbar}{\sqrt{m}}\right)^3\frac{N^{3/2}}{E_{10}^{7/2}},
\end{equation}
\begin{equation}\label{Xmax}
E\equiv E_{10}/E_{20},\ E_{n0}=E_{n}-E_{0}
\end{equation}
and
\begin{equation}\label{Xmax}
X\equiv x_{01}/x_{max}.
\end{equation}
We have defined two scale variables $E$ and $X$, each lying between zero and one, and two scale-free functions $f(E)$ and $G(X)$ whose product can be shown to range between zero and one for the entire range of $(E,X)$.

Eqn \ref{betaMax} is the maximum value of $\beta_{3L}(ext)$. Dividing it into $\beta$ yields the \emph{intrinsic} first hyperpolarizability, a number that is always less than unity and is dimensionless and scale-independent. This is the primary result of the original Letter on the TFL\cite{kuzyk00.01}.

But it is not immediately evident that $\beta_{max}$ is a limit on anything other than $\beta_{3L}(ext)$, the three-level model for which is was calculated.  First, it was calculated using sum rules that are only approximately valid.  Truncation of the sum rules to three levels in order to replace the transition moments in $\beta_{3L}$ with those in Eqns \ref{x02}-\ref{x02xbar22} has, a priori, little justification.  In fact, it is easy to see that sum rule truncation to three levels leads to
\begin{equation}\label{S22}
S_{22}=\hbar^2 N/m\approx -2\left[E_{20}|x_{02}|^2+E_{21}|x_{12}|^2\right]
\end{equation}
which is manifestly violated for any values of the energies and transition moments on the right hand side. The pathologies introduced by working with truncated sum rules to calculate limits was the subject of a rather lengthy paper by Kuzyk\cite{kuzyk14.01}.  An equally insightful work\cite{shafe13.01} showed how it is possible to generate larger maxima in a four state model, and in fact create a catastrophic growth in the maximum in an N-state model as N becomes arbitrarily large.  The N-state catastrophe requires what are likely an unphysical set of spectra and states, but it is not mathematically ruled out.  Given this, what are we to make of the limits discovered using the three-level model?

To make sense of his approach, Kuzyk posited that $\beta\rightarrow\beta_{3L}$ when $\beta$ is near its maximum--that is, at its maximum, only three levels are required to capture most or all of $\beta$.  This is the three-level Ansatz (TLA).  The TLA is not exact, but instead asserts that the maximum hyperpolarizability results when nearly all of the oscillator strength resides in transitions involving the ground state and two excited states.  We know the TLA cannot be exact, because the sum rules are not exact with three levels.  But $\beta$ converges faster than the sum rules, and it is at least believable that $\beta$ might get close to its maximum value using only three levels, while convergence of the sum rules requires many more levels.

We still have a problem, namely to show that $\beta_{3L}\leq\beta_{3L}(ext)$.  If true, then the full sum over states value of $\beta$, calculated with a set $(E_{nm},x_{nm})$ that satisfy the full sum rules approaches $\beta_{3L}$ as $\beta$ approaches its maximum, and this will be bounded by $\beta_{max}$.

But $\beta_{3L}\leq\beta_{3L}(ext)$ is not easily proved, because the spectra and moments used to calculate each factor differ.  $\beta_{3L}$ is calculated using the actual spectra and moments for the problem, while $\beta_{3L}(ext)$ is calculated using approximate moments given in Eqns \ref{x02}-\ref{x02xbar22}.

\begin{figure}\centering
\includegraphics[width=3.4in]{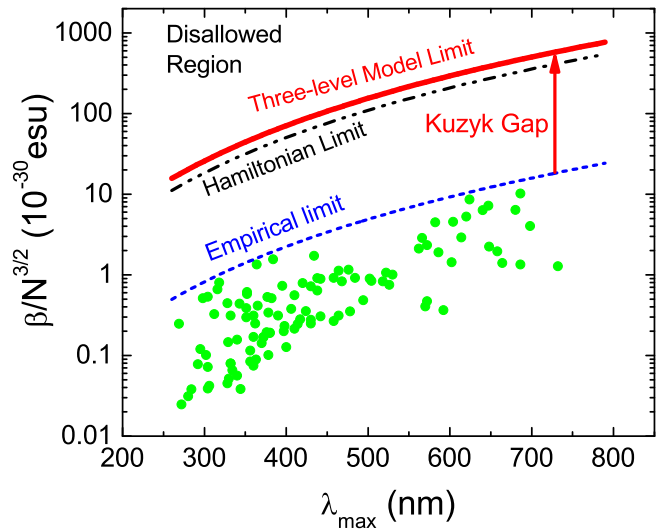}\\
\caption{First hyperpolarizability vs. wavelength of the $E_{10}$ transition.  The (green) data points appearing below the (blue) dashed line labeled \emph{Empirical Boundary} are values measured by EFISH\cite{singe98.01}. The thick solid (red) line labeled \emph{Three-level Model Limit} is that predicted by Eqn \ref{betaMax}. The thinner (black) dot-dash line below it labeled \emph{Hamiltonian Limit} is the highest value calculated directly from many Hamiltonian models.  This figure is a modified version of that first appearing in the original letter\cite{kuzyk00.01,kuzyk03.02}.}\label{fig:betaLimitFig}
\end{figure}

Fortunately, every model system calculated to date for $\beta$ satisfies the TLA and also has $\beta_{3L}\leq\beta_{3L}(ext)$.  Figure \ref{fig:betaLimitFig}, an update of an original plot in Kuyzk's letter\cite{kuzyk00.01,kuzyk03.02} plots the value of $\beta/N^{1.5}$, in esu, against the wavelength associated with the $E_{10}$ level transition.  The dots are experimental EFISH measurements\cite{singe98.01} and represent all of known measurements at that time.  As the Figure implies, the data lie below an apparent or empirical boundary.  The thick solid line labeled \emph{Three-level Model Limit} was published in the (corrected) original Figure\cite{kuzyk03.02} and represents the three-level model prediction of the TFL in Eqn \ref{betaMax}, derived using truncated sum rules.  This was the first published evidence that there may be much more opportunity to synthesize better molecules, if one could learn what properties of the molecules cause their hyperpolarizability to rise toward the limit.

The thin dot-dash line in Figure \ref{fig:betaLimitFig} labeled the \emph{Hamiltonian Limit} is the limit calculated using the Hamiltonian from every model system examined to date.  It falls about $30\%$ short of the Three-level Model Limit.  Since the Hamiltonian is required to generate the sum rules in the first place, the Hamiltonian limit may be the exact limit. The Three-level Model Limit, calculated with a truncated sum over states and truncated sum rules, is approximate at best and may overestimate the actual limit.  A class of unconventional Hamiltonians has not been explored and may generate a maximum response that falls in the limit gap, a topic discussed in Section \ref{sec:issues}.  We can assert that the actual limit lies between the two limit lines depicted in Fig \ref{fig:betaLimitFig}.  With the discovery of a limit, $\beta_{max}$, the first hyperpolarizability tensor $\beta_{ijk}$ may be normalized,
\begin{equation}\label{betaINT}
\beta_{ijk} \rightarrow \frac {\beta_{ijk}} {\beta_{max}}
\end{equation}
to create an intrinsic first hyperpolarizability tensor.  We will often focus on the x-diagonal component of $\beta$ and will assume it has been divided by $\beta_{max}$ to yield an intrinsic first hyperpolarizability $\beta_{int}\leq 1$, by definition.  \textit{In this review, we will use $\beta$ as the intrinsic first hyperpolarizability, going forward}.

In the meantime, the reader's preoccupation might be on Figure \ref{fig:betaLimitFig} and how it is possible that nearly all experimental work published prior to 2000 fell a factor of 30 below the theoretical maximum.  The three-level model can provide some insight into this, as follows.

\begin{figure}\centering
\includegraphics[width=3.4in]{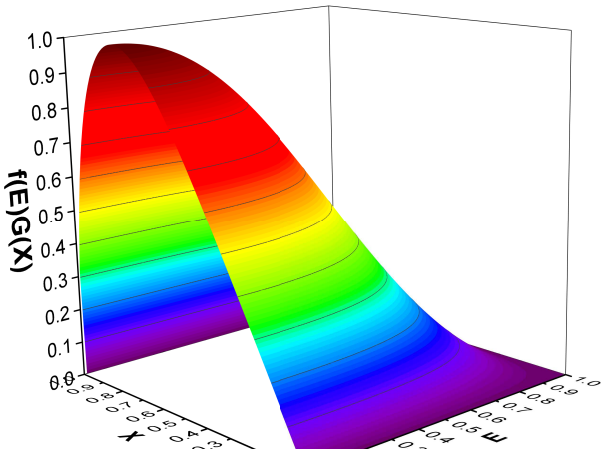}\\
\caption{Surface plot of the dependence of the three-level model $\beta_{3L}(ext)$ on the $E$ and $X$ parameters of this model.  In the figure, $\beta_{3L}(ext)$ is normalized to its maximum value $\beta_{max}$.  The three-level model predicts that the optimum is reached when $X\sim 0.79$ and $E=0$. Coulomb-like molecules have $E$ closer to unity and are predicted by this model to exhibit small intrinsic responses, as is observed.}\label{fig:XEfGsurfaceplot}
\end{figure}

Recall that Eqn \ref{betaMax} was derived by starting with the full sum over states, truncated it to three levels, and invoking constraints from truncated sum rules to eliminate all but the $x_{01}$ transition moment.  This is a true three-level model, because both the SOS and the sum rules are truncated to three levels.  Figure \ref{fig:XEfGsurfaceplot} shows the dependence of the three level $\beta^{3L}_{xxx}(ext)/\beta_{max}=f(E)G(X)$ on the two parameters X and E.  X measures the relative strength of the transition moment between the ground and first excited states, while E measures the ratio of the level difference $E_{10}$ to $E_{20}$.  As $f(E)G(X)$ approaches unity, the scaling parameters take the values $(E,X)\rightarrow (0,0.79)$.  As $E\rightarrow 1$, the model predicts that $\beta \rightarrow 0$. In fact, if $E_{n}\sim 1/n^2$, then the ratio $E\sim 27/32$, suggesting that molecular structures with Coulomb-like potentials should have poor first hyperpolarizabilities, as indeed they all do. Though the model is only approximate when E moves away from zero, it suggests that structures whose energies scale with an inverse power of the state number should have very small $\beta$, while those whose energies scale with some positive power of the state number should have larger $\beta$.  Most structures made to date have a non-ideal energy spectrum\cite{tripa04.01,shafe11.01}, which may in part account for the gap. Ideal spectral scaling is discussed in Section \ref{sec:results}.

\begin{figure}
\includegraphics[width=90mm]{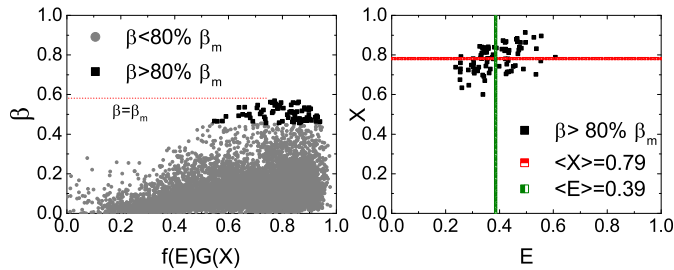}
\caption{Plot of the maximum value of $\beta$ vs $f(E)G(X)$ from a Monte Carlo simulation with thousands of random lengths and orientations of the legs of a nanostar graph (shown in the inset of the right-hand panel).  Near the maximum value of $\beta$, data cluster at the maximum value of the three-level model prediction using the calculated $X$ and $E$ values from the simulation (left). On the right, the parameters $X$ and $E$ converge to universal values when $\beta$ nears its maximum.  (Reprinted from Phys. Rev. A $\bf{87}$, 043824 (2013) with permission from the American Physical Society.)}
\label{fig:3starBetaVsXEfG}
\end{figure}

Despite questions about its accuracy, the factorization of the three-level model into a product of two independent functions of the two scaling variables lends itself to discovering universal properties of molecular and model systems\cite{kuzyk13.01}, in particular specific values for $E$ and $X$ that characterize a topological class of molecular structures.  Figure \ref{fig:3starBetaVsXEfG} illustrates how this is done.  The sum over states analysis for the intrinsic, x-diagonal hyperpolarizability $\beta$ was computed for tens of thousands of shapes of a nanowire \emph{star}, with varying edge lengths and angles relative to one another, and its $X,E$ parameters were calculated\cite{lytel13.01}.  The star nanowire has a maximum intrinsic $\beta$ of about $0.56$, well into the Kuzyk gap, but below the three-level limit of unity. The Figure illustrates the universal scaling property that as the geometry is altered to produce a large response, $X\rightarrow 0.79$, which is what the three-level model would predict at the maximum, but $E\rightarrow 0.39$, well above zero, indicating the spectrum of these star nanowires is suboptimal for $\beta$.  However, the Figure shows exactly how the $X,E$ arrays converge to a universal value that is representative of the star nanowire.  Other structures will have their own scaling limits, but it is nearly always true that the $X$ parameters converge to similar universal values, an indicator that near the maximum of $\beta$, most of the oscillator strength goes into the $0\rightarrow 1$ transition.

The TFL has been extended to include the dispersion of the real and imaginary parts of the first hyperpolarizability\cite{kuzyk06.03} using the dipole-free sum over states method\cite{kuzyk05.02}.  Extensions to include relativistic effects have been derived\cite{leung86.01,cohen98.01,sinky06.01} and used to compute corrections to the nonrelativistic hyperpolarizabilities\cite{dawson15.01}.

A year after the publication of his seminal letter on the TFL for $\beta$, Kuzyk published his letter showing how to compute a maximum and minimum value of $\gamma$ using truncated sum rules, again in a three-level model\cite{kuzyk00.02}. The second hyperpolarizability tensor $\gamma_{ijkl}$ is a fourth rank tensor and may be calculated with a sum over states\cite{orr71.01}:
\begin{eqnarray}\label{gammaijkl}
\gamma_{ijkl} &=&  \frac{1}{6} P_{ijkl}  \left({\sum_{n,m,l}}' \frac{r_{0n}^{i} \bar{r}_{nm}^{j}\bar{r}_{ml}^{k}r_{l0}^{l}}{E_{n0} E_{m0} E_{l0}} \right. \nonumber  \\
&& \left. - {\sum_{n,m}}' \frac{r_{0n}^{i}r_{n0}^{j}r_{0m}^{k}r_{m0}^{l}}{E_{n0}^2 E_{m0}}\right),
\end{eqnarray}
where the permutation operator here is over the four indices $(i,j,k,l)$.  Applying the three-level model to $\gamma$, we find a maximum given by
\begin{equation}\label{gammaMax}
\gamma_{max} = 4 \left(\frac{e\hbar}{\sqrt{m}}\right)^4 \frac{N^{2}}{E_{10}^{5}} .
\end{equation}
and a minimum equal to $-0.25$ of the maximum.  It is worth noting that Kuzyk presented these results in his original letter\cite{kuzyk00.01}, but his later letter\cite{kuzyk00.02} explicitly derived them.

As he did for $\beta$, Kuzyk defines an intrinsic second hyperpolarizability
\begin{equation}\label{gammaINT}
\gamma_{ijkl} \rightarrow \frac {\gamma_{ijkl}} {\gamma_{max}}.
\end{equation}
Both intrinsic hyperpolarizabilities in Eqns \ref{betaINT} and \ref{gammaINT} have the property that they are the same for all structures of the same shape but with scaled lengths, $i.e.$ for all structures whose lengths are rescaled according to $a_i \rightarrow \epsilon a_i$.  Artifacts due to simple size effects are eliminated by using these intensive quantities. As with $\beta$, \textit{all second hyperpolarizabilities discussed beyond this point are labeled $\gamma$ and are intrinsic values}.  Eqns \ref{betaMax} and \ref{gammaMax} are the fundamental results of the TFL and place absolute bounds on the size of optical nonlinearities in any structure, independent of the underlying physics of the structure.  For reference, Table \ref{tab:modelLimits} summarizes the exact limits corresponding to the two limit lines in Figure \ref{fig:betaLimitFig}.

\begin{table}\small\centering
\caption{Limit predictions for the intrinsic, off-resonant hyperpolarizabilities by the three-level model and from all Hamiltonian models studied to date.}\label{tab:modelLimits}
\newcolumntype{S}{>{\arraybackslash} m{3cm} } 
\begin{tabular}{S c c c }
\\
  \hline\hline
Model  & $\left|\beta_{max}\right|$ & $\gamma_{max}$ & $\gamma_{min}$ \\
  \hline\hline
Three-level model & 1.0 & 1.0 & -0.25 \\
Hamiltonian         & 0.7089 & 0.6 & -0.15 \\
\hline\hline
\\
\end{tabular}
\end{table}

An extensive review of the application of sum rules and scaling in nonlinear optics has been produced in Physics Reports\cite{kuzyk13.01}. Sum rules are an invaluable tool for understanding physics of quantum systems.  The literature is replete with earlier applications, including sum rules for particle physics\cite{jacki67.01}, TRK generalizations\cite{wang99.01}, dispersion relations in nonlinear optics\cite{bassa91.01}, nonlinear sum rules\cite{scand92.01}, and summation of series\cite{mavro93.01}.

\subsection{Post-TFL experimental research}\label{sec:postTFL}
Kuzyk's letter on the TFL in 2000 seems to have attracted minimal attention during the first year after its publication.  Today, it is one of the most cited references in nonlinear optics and has gained general acceptance among researchers working in the field.  But given the size of the gap between theory and experiment at the time of publication of the Letter, a number of researchers caught on to its significance and drove developments in the years immediately following publication. Some of the salient ones are summarized here.

Shortly after the publication of the first Letter on the TFL, Clays\cite{clays01.01} showed that the TFL applied to the theoretical prediction of multiphoton fluorescence, and then applied the TFL to analyze two design strategies for engineering large second-order nonlinear optical responses\cite{clays03.01}.  Kuzyk applied the TFL to understand the limits on two-photon absorption cross sections\cite{kuzyk03.03}.  The work was extended to include all second order nonlinear optical phenomena\cite{kuzyk06.03}.

Tripathy et al. employed linear and Raman spectroscopy, Hyper-Rayleigh scattering, and Stark spectroscopy to determine relevant parameters contributing to the nonlinearities\cite{tripa04.01}.  Included were tests of dilution effects due to vibronic states, investigations of unfavorable energy spacing of the molecule or atom, simplifications inherent in local field models, and an analysis of the effects of truncation of the sum rules. These studies concluded that the energy spectrum of real systems compared with the ideal is the most likely factor that keeps the hyperpolarizabilities of real molecules well below the limit.

Slepkov et al. synthesized a series of polyynes with up to 20 consecutive sp-hybridized carbons, measured their non-resonant $\gamma$ values, and developed a scaling law for $\gamma$ as a function of the number of acetylene repeat units\cite{slepk04.01}.  Their scaling exponent compared favorably with that predicted by the TFL.

In what must have been anticipation of forthcoming theoretical explorations of the fundamental limits with potential and statistical models, Kuzyk invented the dipole-free sum over states (DFSOS)\cite{kuzyk05.02}, an ingenious tool that eliminated dipole moment differences, allowing the use of \emph{diagonal} sum rule constraints.  The DFSOS was subsequently explored by Champagne et al. for push-pull $\pi$-conjugated systems\cite{champ06.01} and was adopted in the following years for Monte Carlo calculations of the hyperpolarizabilities\cite{kuzyk08.01,shafe10.01}.

In a 2006 paper, Zhou et al. discovered that a conjugated bridge between donor-acceptor molecules, with many sites of reduced conjugation to impart a modulated conjugation path for electrons, led to a first hyperpolarizability that fell well into the gap and within $30\%$ of the fundamental limit\cite{zhou06.01}.  This work numerically calculated $\beta$, starting with a hyperbolic tangent potential, then used an optimization algorithm to continuously vary the potential to maximize $\beta$. They achieved a maximum $\beta$ near the Hamiltonian limit in Table \ref{tab:modelLimits}, and universal scaling parameters $E\sim 0.314$ and $X\sim 0.775$, corresponding to a three-level model $\beta_{ext}=0.89$, about $18\%$ above the actual $\beta$. The difference is once again due to the truncated sum rules used in the three-level model, but not in the exact calculation. In this model, their first and sixth excited states were the only two with appreciable overlap with the ground state, showing that the three-level Ansatz was in force.

Shortly thereafter, Perez-Moreno et al. published a letter demonstrating a series of donor-acceptor chromophores with a modulated conjugation path between them\cite{perez07.01}.  This approach modulated the amount of aromatic stabilization energy along the conjugated bridge, inducing the modulation through use of aromatic moieties with varying degrees of aromaticity.  Hyper-Rayleigh scattering was used to measure an enhanced intrinsic first hyperpolarizability of nearly $0.05$, or only twenty times below the limit\cite{perez07.02}, and the early results seemed to confirm that modulated conjugation would lead to enhancement of $\beta$\cite{perez09.01}.

May et al. investigated the second hyperpolarizabilities of a class of donor-substituted cyanoethynylethene molecules, off-resonance\cite{may05.01}.  The molecules showed a high efficiency due to a two-dimensional conjugated system and effective donor-acceptor substitution patterns.  And in 2007, May et al. studied extended conjugation and donor-acceptor substitution to enhance $\gamma$ in small molecules, again using the TFL to discover a weak power-law dependence for $\gamma$ that depended on the number of conjugated electrons separating the donor and acceptor, and asserting that its origin was the competition between the energy separation $E_{10}$ and the strength of the transition moments, both of which depend on the conjugation length\cite{may07.01}.  They achieved highly efficient molecules within a factor of 50 of the theoretical limit for centrosymmetric molecules.

The first and second order response of twisted $\pi$-electron chromophores were identified to constitute a new paradigm for enhanced electro-optic materials\cite{kang05.01,kang07.01,he11.01} based on the fact that they fell far into the gap between the best molecules ever measured and the fundamental limit\cite{zhou08.01}.

Analysis of experimental data on certain dye molecules using a combination of measurements and sum rules led to an accurate prediction of the imaginary part of the second hyperpolarizability\cite{perez11.02}, which is only nonzero when incorporating linewidths of the contributing levels. The major advance was the achievement of a reduction in the number of physical measurements required by using a combination of measurement techniques and the constraints imposed by the self-consistency of the TRK sum rules.

More recently, Jiang et al. have designed coupled porphyrin chromophores with large off-resonance $\beta$ using density functional and coupled-perturbed Hartree-Fock computations\cite{jiang12.01}.  Van Cleuvenberger et al. synthesized di-substituted poly(phenanthrene), a conjugated polymer, and studied it with hyper-Rayleigh scattering\cite{cleuv14.01}.  Most interesting is that this compound showed one of the highest $\beta$ measured, despite the absence of a donor-acceptor motif.  The authors attribute the large response to a modulation of the conjugation along the polymer backbone. Al-Yasari et al. have characterized donor-acceptor organo-imido polyoxometalates with static first hyperpolarizabilities that exceed those of comparable compounds\cite{al2016donor}.  And Coe et al. have investigated the second-order nonlinear optical response of Helquat dyes with large static first hyperpolarizabilities\cite{coe2016helquat}.

Every reference just cited employed the TFL to understand and interpret their calculations and measurements.  But it is not clear which approach of the many described represents a general design principle for nonlinear optical molecules, though the modulation of the conjugation path between donor-acceptor groups (and even without donor-acceptors) appears to offer a quantitative explanation of the enhancement and of the validity of the three-level Ansatz.  The world of synthetic and computational chemists and physicists has certainly caught on to the concept of the TFL and sought methods to achieve them.

In April of this year, two preprints appeared which analyze classes of molecules according to their universal scaling properties, one paper each for $\beta$\cite{perez16.01} and $\gamma$\cite{perez16.02}. The authors' stated objective was to identify classes of molecules with superior scaling properties as they increased in size. Naive scaling of a molecule with length and electron count is that predicted on dimensional grounds and from the sum over states, which shows a $\beta_{max}\sim N^{1.5}/E_{10}^{3.5}$ where N is the electron count and $E_{10}$ is the ground to first excited state energy level difference.  Dividing $\beta$ of a structure by $\beta_{max}$ eliminates the dimensional dependence, leaving a scalar number characterizing the molecule.  But if the molecule is part of a class whose length increases as additional bonds are added, $\beta$ may scale slower (sub-scaling), the same (nominal scaling), or faster (super-scaling) in the nomenclature of the authors.  The authors discovered that some molecular structures make sub-optimal use of additional electrons as they grow in length, while others have higher efficiency and produce larger response than predicted by naive scaling with $N$ and $E_{10}$.  This result holds for both the first and second hyperpolarizabilities.  The papers offer a set of design paradigms for determining which molecular units should be linked to provide enhanced nonlinear optical responses.  Both papers are included in the special issue in which this review is to be published.  The papers include tabulations of the results for many of the systems described in this section and are an up to date summary of the post-TFL efforts to develop more active molecules that exceed the empirical limit in Figure \ref{fig:betaLimitFig}.

\section{Theoretical explorations of the limits}\label{sec:results}
The nonlinear optical hyperpolarizability tensors $\beta_{ijk}$ and $\gamma_{ijkl}$ of any structure may be calculated from its Hamiltonian by perturbation theory using a sum over states\cite{orr71.01}, Dalgarno-Lewis quadrature\cite{dalga55.01}, and by other quadrature methods. In general, the topological properties of the structure are reflected in the spectra, while the geometrical properties--symmetries and shapes--are manifest in the transition moments.  If the Hamiltonian is parametrized and then diagonalized, sets of spectra and transition moments $(E_{nm},x_{nm})$ that are functions of the parametrization may be used in a sum over states to parametrically study the dependence of the hyperpolarizability tensors on features that describe the molecule that generated them.  Optimization of the hyperpolarizabilities can, in principle, determine which features of a molecule are most critical in maximizing its nonlinear optical response.  All such calculations have at best reached the Hamiltonian Limits in Table \ref{tab:modelLimits}.

Explorations of the kinds of systems that might lead to responses at the limits in Table \ref{tab:modelLimits} are often optimization computations using parametrized potential models not necessarily representing a specific molecule structure, and have been studied to determine the maximum range of the hyperpolarizabilities and its dependence upon the parameters describing the shape of the potential, electron interactions, the dimensionality of the system, and the distribution of charge. Most of these potential optimization models skip the direct computation of the sets of spectra and moments, and obtain the hyperpolarizabilities through quadrature or differentiation of appropriate functions of the solutions to the potential model. In these models, the TRK sum rules hold exactly, so long as the Hamiltonian satisfies the required double commutator with the position operator.  All of these computations to date have also achieved at best the Hamiltonian limits in Table \ref{tab:modelLimits}.

Other exploration methods start with the sets $(E_{nm},x_{nm})$ and select them at random, but use the TRK sum rules to constrain them.  These Monte Carlo models compute the hyperpolarizabilities using a sum over states, typically the dipole-free form\cite{kuzyk05.02}.  Since the computations start with the spectra and moments, and not the Hamiltonian, the results may not even correspond to a physical Hamiltonian.  Even if all possible points in the $(E_{nm},x_{nm})$ parameter space are sampled, this set may have larger cardinality than those sets of spectra and moments arising from any physical Hamiltonian.  In this sense, the Monte Carlo models might be making predictions that are non-physical.  In all Monte Carlo simulations to date, the maximum values of the hyperpolarizabilities are the three-level model limits in Table \ref{tab:modelLimits}.

Many of the analytical explorations of the fundamental limits have used one-dimensional models to explore a single diagonal tensor component.  These models can study range of limits possible in systems, but not their dependence on symmetry, geometry, or topology. Notable exceptions are the studies of the effects of geometry on the hyperpolarizability in systems with a two-dimensional Coulomb potential\cite{kuzyk06.02,zhou07.02}.  Model calculations have also examined the effect of electron interactions using one-dimensional potentials\cite{watki11.01} and quasi-one dimensional quantum graphs\cite{lytel15.02}

The rest of this section highlights the key results from potential optimization, Monte Carlo simulations, and quantum graph models.  The section closes with a quantitative example of the three-level Ansatz (TLA) and its emergence from the physics when the hyperpolarizabilities approach their maximum values.

\subsection{Hamiltonian models.}\label{sec:Ham}
The problem of determining which set of spectra and eigenstates of a system will maximize the intrinsic nonlinear optical response is essentially intractable.  It is impossible to write down a general Hamiltonian, solve it, and calculate the nonlinearities.  Nevertheless, the effects of several different parameters, including molecular geometry, external electromagnetic fields, and electron-electron interactions, have been investigated. The effect of molecular geometry on the hyperpolarizability is determined by varying the positions and magnitudes of charges in two dimensions and correlating them with dipolar charge asymmetry and the variations of angle between point charges in octupolar structures. It was shown that the best dipolar and octupole-like molecules have intrinsic hyperpolarizabilities near 0.7\cite{kuzyk06.02}.

One of the earliest potential models studied for nonlinear optics is the clipped Harmonic oscillator, defined on an infinite half-space\cite{tripa04.01}.  For this system the maximum value of the three level model $\beta$ is of order $0.6$, and the full value computed from the sum over states is about $0.57$.  This result was significant in that it was one of the first to verify that the empirical limit in Figure \ref{fig:betaLimitFig} was not a fundamental limit.

The first optimization centered on finding a one-dimensional potential that could generate a $\beta$ near the Hamiltonian limit used a Nelder-Mead simplex algorithm to optimize the dipole-free sum over states\cite{zhou06.01}.  The initial potential was a hyperbolic tangent on a one-dimensional line bounded by infinite potential at either end. The algorithm continuously varied the potential to maximize $\beta$ at each step and testing for convergence.  For each potential, a quadratic finite element method was used to compute the wavefunctions and energy levels.  The procedure showed that a modulated potential across a simulated donor-acceptor bridge produced a maximum $\beta$ equal to the Hamiltonian limit shown in Table \ref{tab:modelLimits}, and revealed that at the optimum value, only three levels had appreciable spatial overlap to contribute to the sum over states.  In a simple calculation, the analysis showed that the Hamiltonian limit was achievable in a one-dimensional potential model and that the three-level Ansatz was operative.  But the model also revealed that the three-level model parameters were $E=0.314$ and $X=0.775$, leading to $f(E)G(X)=0.89$, an overestimate of the actual value computed when all sum rules are valid.  Most important, the results stimulated experimental synthesis of a modulated conjugation donor-acceptor system with enhanced response\cite{perez07.01}, showing for the first time hints of how optimizations in the TFL could lead to practical design rules for new molecules.  As noted in Section \ref{sec:postTFL}, the modulation was induced using aromatic moieties with different degrees of aromaticity comprising the asymmetrically substituted $\pi$ bridge.

At about the same time, the effect of geometry on the hyperpolarizability was initially studied using two-dimensional logarithmic potential functions simulating superpositions of force centers representing nuclear charges lying in various planar geometrical arrangements\cite{kuzyk06.02}.  The eigenvalue problem was solved using a quadratic finite element method and generated the lowest few dozen states and spectra for use in a sum over states computation of $\beta$.  The Hamiltonian limits in Table \ref{tab:modelLimits} were achieved for certain octupolar-like structures with donors and acceptors of varying strength at the branches.  Once again, the three-level Ansatz was valid.

It is essential to emphasize that the predictions of these early studies were by no means a foregone conclusion.  Up to this point, the fact of a factor of 30 gap between experiment and theory was still a surprise and not well understood, and the validity of the TFL described in Section \ref{sec:history} still needed to be established.

The optimization of potentials in one dimension was extended to study a series of starting potentials, including the hyperbolic tangent, linear, quadratic, square root, and sinusoidal with a linear growth\cite{zhou07.02}.  The authors abandoned the sum over states method used in their prior optimization paper\cite{zhou06.01} for a finite field approach which calculated first the ground state of the potential model in the presence of a static electric field and computed $\beta$ by differentiating twice with respect to the field.  In all cases, optimization led to an upper limit on $\beta$ equal to the Hamiltonian Limit in Table \ref{tab:modelLimits}.  This optimization hinted that the achievement of large nonlinearities may be insensitive to the exact form of the potential.  More important, the analysis showed that the sum over states expression, truncated to N levels, had more adjustable parameters as N increased, raising the question of how the three-level Ansatz, which held for each optimized computation, emerged as an accurate description of the physics. This question remains unanswered today and is discussed in Section \ref{sec:issues}.  The authors then used their finite field algorithm to optimize $\beta$ for the two dimensional logarithmic superposition potential\cite{kuzyk06.02} and, curiously, obtained an upper limit of $0.685$, just below the Hamiltonian limit.  This lower value may suggest that the limits in Table \ref{tab:modelLimits} depend on the dimension of the structure.  This remains to be explored in multidimensional optimization calculations.

Further optimizations were explored using external electromagnetic fields and nuclear placement\cite{watki09.01}, and the effect of electron interactions was analyzed and showed the same universal properties as the one electron systems\cite{watki11.01}.  This was the first evidence that the near-limit hyperpolarizabilities were not an artifact of studying one electron systems.  At this point, the literature showed that optimized potentials, with interacting electrons and in the presence of electromagnetic fields, could achieve the Hamiltonian limits, and when they did, the three-level Ansatz applied.  Most important, it appeared that this conclusion was insensitive to the exact form of the potential.  Modulated conjugation, a prediction of an early optimization model, was verified experimentally to yield an enhancement which remained well below the Hamiltonian Limit\cite{perez07.01,perez09.01}.  As of yet, no general design rules had emerged for real molecules that might achieve a sizable fraction of those limits.

In a further effort to understand which parameters and how many are essential in a potential to maximize either $\beta$ or $\gamma$, two recent papers\cite{ather12.01,burke13.01} employed an innovative optimization algorithm with minimally parametrized potentials and were able to achieve the Hamiltonian limits for both hyperpolarizabilities, but with a poor definition of the potential function.  The analysis strove to determine how sensitive the optimizations were to the parametrization details of the potential, by using the Hessian matrix, which has eigenvalues that are essentially the local principal curvature of the objective function at the maximum.  This is effectively an effort to construct the potential from the optimized hyperpolarizabilities, a sort of inverse problem to the usual optimization approaches, though the authors did assume the potential was a piecewise linear function with many segments.  The analysis for $\beta$ achieved the Hamiltonian limit with only two parameters, a result that is not going to help much in determining the ideal molecular potential.  The Hamiltonian limit for $\gamma$ was achieved with a single parameter. The authors concluded that numerically optimized potentials need careful interpretation before positing them as design paradigms.

Recently, the piecewise potential optimization approach was applied to maximize the hyperpolarizability in one dimension of a multiple electron system\cite{burke16.01}.  The potential models included a $\delta$ potential, a sort of triangular potential, and the piecewise potential used in the authors' earlier studies\cite{ather12.01,burke13.01}.  These results are quite interesting, because the authors found $\beta\rightarrow 0.40$, $\gamma^{+}\rightarrow 0.16$, and $\gamma^{-}\rightarrow -0.061$ for $N\geq 8$ electrons.  These bounds were achieved with only two parameters, and no improvements were possible with more general potentials. The results are interesting because they show a reduction in the maximum hyperpolarizability with many non-interacting electrons, in contrast to the two interacting electron work previously cited\cite{watki11.01}.  More research is indicated to further explore the significance.

\subsection{Monte Carlo simulations}\label{sec:monte}
An entirely different approach to discovering the nature of the fundamental limits and the systems that attain them emerged from a series of numerical simulation studies that choose sets of spectra and transition moments at random, but constrain them to satisfy the sum rules.  The method uses the dipole-free sum over states\cite{kuzyk05.02} to eliminate dipole moment differences from the usual sum over states, thus enabling diagonal sum rules to be used as constraints to compute a set of off-diagonal transition moments for use in a large-scale numerical simulation of $\beta$ and $\gamma$.  A series of papers\cite{kuzyk08.01,shafe10.01}, on Monte Carlo simulations of nonlinear optical systems whose spectra and moments are randomly selected but constrained by the TRK sum rules revealed a maximum value of unity for $\beta$ and $\gamma$, and a minimum value of negative one quarter for $\gamma$, as predicted by the TFL.  These simulations showed that the predictions of the TFL were attainable by \emph{some} set of energies and eigenstates, but with no insight into what physical system, if any, could generate these parameters.

\begin{figure}\centering
\includegraphics[width=3.4in]{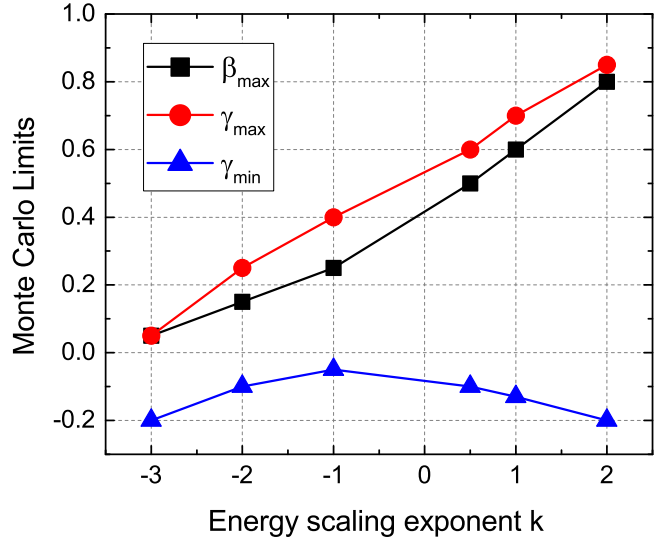}\\
\caption{Spectral scaling of the largest values of the hyperpolarizabilities with energy scaling exponent k, where the energy as a function of eigenmode number $n$ takes the form $E_{n}\sim n^{k}$ . The data are the results of a Monte Carlo calculation\cite{shafe11.01} and show the mode-scaling behavior of both hyperpolarizabilities.  Note that the maximum values shown exceed the Hamiltonian Limits and are bounded by the Three-level Model Limits.}\label{fig:MCbetaGammaEscaling}
\end{figure}

Monte Carlo studies of large numbers of constrained but otherwise random sets of moments, and spectra scaling with the eigenstate number to some power revealed that the optimum spectrum for a system scales quadratically or faster with eigenstate number \cite{shafe11.01}.  Figure \ref{fig:MCbetaGammaEscaling} shows how the maximum intrinsic hyperpolarizabilities scale with the energy scaling exponent k, defined through $E_{n}\sim n^{k}$.  Topological properties of a system determine the spectra, while geometrical properties determine projections of the transition moments onto a specific external axis.  An optimum topology is one which produces a spectrum scaling quadratically or faster.  As noted in Section \ref{sec:history} using the  three-level model, Coulomb-like potentials generate spectra which scale inversely as some power of the mode number, and these fall far below the optimum.  The Monte Carlo work suggests this is the primary reason for the Kuzyk gap in the first place.

The Monte Carlo simulations are the only models which reach the three-level model limits in Table \ref{tab:modelLimits}.  Since these limits are higher than the Hamltonian Limits, the TFL has a \emph{limit} gap whose origin remains to be explained.  This is discussed further in Section \ref{sec:issues}.

\subsection{Quantum graph models}\label{Qgraphs}
Quantum graphs are quasi-one dimensional systems to which electron dynamics are confined along the edge of a network structure which can be thought of as a branched nano-wire structure, or a quasi-linear molecule, such as a donor-acceptor, with side groups.  Quantum graphs were first studied as tractable molecular models\cite{pauli36.01,kuhn48.01,ruede53.01,scher53.01,platt53.01} and have been invoked as models of mesoscopic systems\cite{kowal90.01}, optical waveguides \cite{flesi87.01}, quantum wires\cite{ivche98.01,sanch98.01}, excitations in fractals \cite{avish92.01}, and fullerines, graphene, and carbon nanotubes\cite{amovi04.01,leys04.01,kuchm07.01}. Quantum graphs are also exactly solvable models of quantum chaos\cite{kotto97.01,kotto99.02,kotto00.01,blume01.04}.  Microwave networks have recently been successfully used to experimentally simulate quantum graphs\cite{hul04.01}.

The nonlinear optics of electron confinement to a graph edge was first explored in 2011\cite{shafe11.02}.  This launched the injection of quantum graphs into nonlinear optics\cite{shafe12.01,lytel12.01,lytel13.01,lytel13.03,lytel13.04,lytel14.01,lytel15.01} as model systems exhibiting the optimum spectra for large intrinsic nonlinear optical response for specific geometries.  They are a general tool for studying the dependence of the hyperpolarizabilities on the geometry of a structure, and when the Cartesian tensors are resolved into the spherical tensor basis, spatial distributions of large numbers of simulations easily reveal symmetries that will maximize the nonlinear optical response\cite{lytel13.01}.

Quantum graphs are ideal tools for exploring the nature of the fundamental limits of Hamiltonian systems.  Their energy spectra are functions of the edge lengths and angles relative to a fixed axis, as well as the number of star vertices and potentials operating on the edges.  They are quasi-quadratic, in general, with values lying between root boundaries that scale quadratically with mode number, and as such, they should exhibit some of the largest responses possible.  Using Monte Carlo methods, dozens of topologies with tens of thousands of geometries for each have been explored to determine the ranges of hyperpolarizabilities for various structures.

\begin{figure}\centering
\includegraphics[width=1.7in]{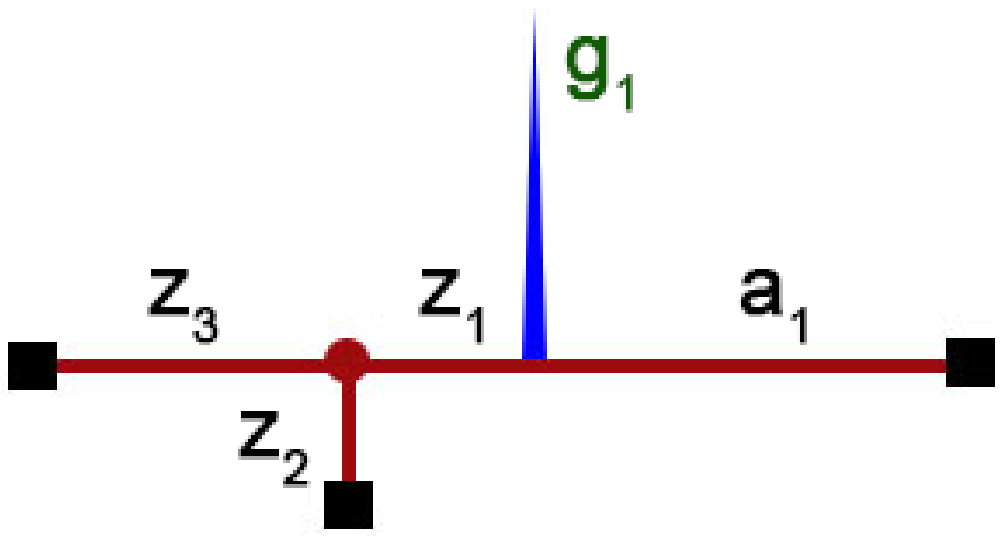}\includegraphics[width=1.7in]{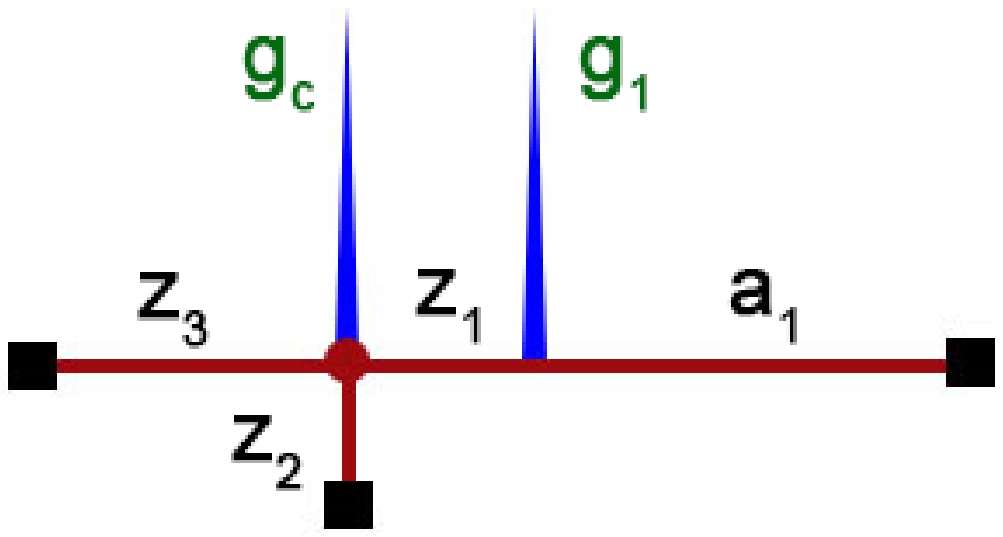}
\caption{Quantum graph with one prong and one $\delta$ function located on a main edge of length $z_3+z_1+a_1$, left panel.  A similar graph but with two $\delta$ functions, one located at the prong position, is shown on the right.  The hyperpolarizabilities of this class of graphs are displayed in Table \ref{tab:QGresults}.}\label{fig:graphs}
\end{figure}

Figure \ref{fig:graphs} depicts two quantum graphs from a class of graphs with $\delta$ functions and \emph{prongs} located at various positions on a straight edge (wire)\cite{lytel15.01}.  The presence of the $\delta$ functions or prongs creates a giant enhancement of the nonlinear optical response due to a effect known as \emph{phase disruption} of the lowest eigenfunctions of the graph\cite{lytel15.02}.  This class of graphs has the optimum topology for nonlinear optics\cite{lytel13.04,lytel14.01,lytel15.01}.

Table \ref{tab:QGresults} presents a compilation of results for the class of graphs depicted in Figure \ref{fig:graphs}.  The results indicate that the Hamiltonian Limits apply to quantum graphs.  None were observed to have a hyperpolarizability that exceeded these limits.  The results are consistent with all prior Hamiltonian models and show that neither geometry, nor topology will bridge the new gap between Hamiltonian and Three-Level model limits.
\begin{table}\small\centering
\caption{Effects of geometry and topology in the optimum quantum graphs ($\delta$ function graphs, the prong graphs, and hybrids of these, as explained in the text).  The number of $\delta$s and prongs for each are displayed in the table.  $\beta$ is the value of the first hyperpolarizability's x-diagonal component when the graph is rotated to the position maximizing it.  The minimum value is just the negative of the maximum value.  $\gamma^{+}$ is the value of the second hyperpolarizability's x-diagonal component when the graph is rotated to the position maximizing it. ($\gamma^{-}$ is the minimum value upon rotation\cite{lytel13.04}.}\label{tab:QGresults}
\newcolumntype{S}{>{\arraybackslash} m{2.8cm} } 
\begin{tabular}{S c c c c c }
\\
  \hline\hline
Topology  & prongs & $\delta$ & $\beta$ & $\gamma^{+}$ & $\gamma^{-}$ \\
  \hline\hline
Bare                          & 0 & 0 & 0 & 0 & -0.126 \\
One $\delta$                  & 0 & 1 & 0.680 & 0.580 & -0.126 \\
Two $\delta$                  & 0 & 2 & 0.697 & 0.588 & -0.126 \\
Three $\delta$                & 0 & 3 & 0.709 & 0.590 & -0.126 \\
One prong                     & 1 & 0 & 0.573 & 0.296 & -0.126 \\
Two prongs                    & 2 & 0 & 0.610 & 0.353 & -0.126 \\
One $\delta$ at prong         & 1 & 1 & 0.644 & 0.370 & -0.133 \\
One $\delta$ not at prong     & 1 & 1 & 0.692 & 0.378 & -0.145 \\
Two $\delta$ at/not at prong  & 1 & 2 & 0.684 & 0.583 & -0.147 \\
\hline\hline
\\
\end{tabular}
\end{table}

A key observation from Table \ref{tab:QGresults} is that the compressed $\delta$ atom, a quantum wire with a single $\delta$ function and infinite potential at either end, has nonlinearities that approach the Hamiltonian Limits, and that the addition of only one or two more $\delta$ functions is sufficient to reach the Hamiltonian Limits.  This result corroborates the linear piecewise potential optimization results discussed in an earlier subsection\cite{ather12.01,burke13.01}.  A new design concept, \emph{phase disruption}, was posited to explain these results and provide insight into a possible new design paradigm\cite{lytel15.02}.

A multi-electron version of the quantum graph model was developed to confirm that graphs with $N$ electrons could still achieve a sizable fraction of the Hamiltonian limits\cite{lytel15.02}. The limits were similar to those recently reported by Burke et al.\cite{burke16.01}.  In each model, the electrons were non-interacting, however, but models accounted for the Fermion properties of electrons in assembling the eigenstates.

\subsection{Three-levels for $\beta$, four for $\gamma$}\label{sec:3and4levels}
The three-level Ansatz, introduced in Section \ref{sec:history}, posits that only three states are required to compute $\beta$ when it is near the limit.  For $\gamma$, four states are required. The origin of this result is unknown, and may be unprovable\cite{shafe13.01}, but it stands as an empirical result for all computational models and experiments to date.  This is a rather profound observation, not dissimilar to the Godel theorem about provability in arithmatic systems.

To a casual observer, the TLA may seem an obvious result: The three level model $\beta(ext)=f(E)G(X)$ of Section \ref{sec:history} holds when only three states contribute to $\beta$ and the sum rules have been truncated to three levels.  [Reminder to reader: $\beta$ has been normalized to $\beta_{max}$]. Under these conditions, two parameters, E and X, are sufficient to describe variation of the nonlinearity with variations in the spectra and transition moments.  Though it may seem an accurate approximation, especially with the validity of the TLA, it fails precisely because the sum rules are manifestly invalid under these circumstances.  It remains a useful model for scoping systems by their E and X values, but is approximate.  So it is not clear why the TLA should be valid.  But without its validity, the computation of $\beta$ with an N-state model leads directly to the N-state catastrophe\cite{shafe13.01}.  The TLA is valid, but we don't know why or how.  Proving its validity remains one of the outstanding theoretical problems in the TFL--see Section \ref{sec:issues}.

Allowing the sum rules full validity by including all states and spectra, but restricting the computation of $\beta$ to three levels, gives a value we called $\beta_{3L}$ in Eqn \ref{beta3L}. It is this value of $\beta$ to which the exact value converges as $\beta$ approaches the limit.  Here, the TLA is consistent with the convergence of the sum over states $\beta$ to one constructed using only three levels in Eqn \ref{beta3L}.  The fact that this convergence occurs can be attributed to the TLA, if one likes, but the converse is equally true, and neither observation can be used to prove the other.
\begin{figure}\centering
\includegraphics[width=3.3in]{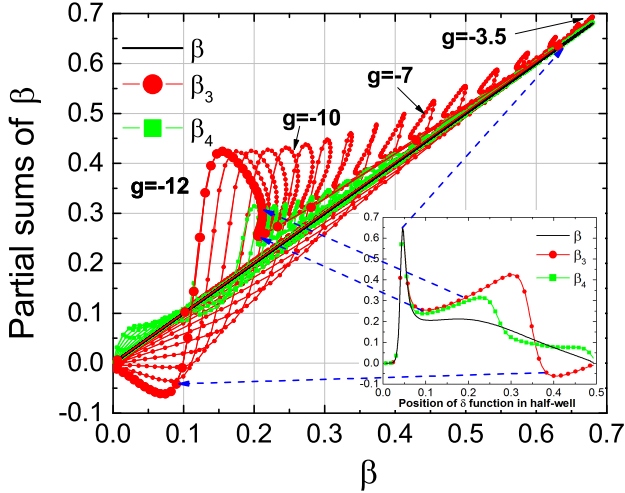}\\
\caption{The three- and four-level partial sum over states for $\beta$ for a dressed $\delta$ atom, plotted against $\beta$ for the full sum over states.  The dots on each arc are the values obtained for a fixed potential strength when the $\delta$ function is moved from left to center of the structure. The many arcs correspond to many values of $g$, with the maximum $\beta$ occurring when $g\sim -3.5$.  The three-level sums converge faster to the actual value for this topology as their value of $g$ approaches the optimum.  The four-level sums nearly always converge to the correct $\beta$ (see text).}\label{fig:1delta_beta_TLA}
\end{figure}

\begin{figure}\centering
\includegraphics[width=3.3in]{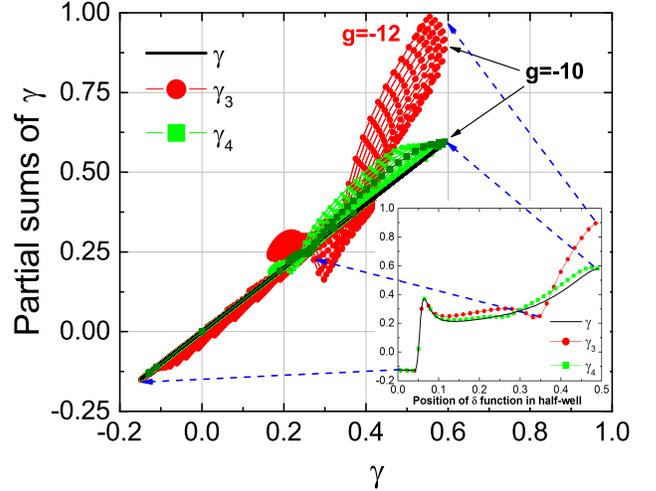}\\
\caption{The three- and four-level partial sum over states for $\gamma$ for a dressed $\delta$ atom, plotted against $\gamma$ for the full sum over states.  The dots on each arc are the values obtained for a fixed potential strength when the $\delta$ function is moved from left to center of the structure. The many arcs correspond to many values of $g$, with the maximum $\beta$ occurring when $g\sim -10$.  The three-level sums do not converge to the actual value for this topology as their value of $g$ approaches the optimum, but the four-level sums nearly always converge to the correct $\gamma$ (see text).}\label{fig:1delta_gamma_TLA}
\end{figure}

Still, the convergence is rather staggering, as illustrated in Figure \ref{fig:1delta_beta_TLA} for $\beta$ and Figure \ref{fig:1delta_gamma_TLA} for $\gamma$.  Each plots the partial sums for three and four levels for each of $\beta$ and $\gamma$, respectively, against the sum for each with all states, for a compressed $\delta$ atom\cite{blume06.01}, also known as a one-$\delta$ quantum graph\cite{lytel13.04}. This system is simply a bare wire, or one-dimensional quantum well, with a single, finite $\delta$ function potential of strength $g/2L$ located somewhere in the well.  For $g=0$, we find $\beta=0$ by centrosymmetry, and $\gamma=-0.126$.  Addition of a single $\delta$ function drives the maximum of both $\beta$ and $\gamma$ to values near the Hamiltonian Limits, an observation to which we return in Section \ref{sec:issues}.  The Figures are the result of a series of calculations via the sum over states, where the strength $g$ is stepped in increments of $0.5$ from $-12$ to higher values, and with the position of the $\delta$ function stepped from one end to the other.  Each set of closely spaced dots represents a value of $\beta$ (or $\gamma$) for a fixed value of $g$ as the position from the left wall of the well is varied from zero to the midpoint of the wire, as indicated in the insets.
We have by no means proved that all systems will asymptote to the Hamiltonian limits in Table \ref{tab:modelLimits}.  But we can see how a simple potential model can approach these limits and how the three-level Ansatz operates for $\beta$, with an extra level required for $\gamma$.  To date, there are no theoretical or experimental exceptions to this behavior.

\section{Open questions}\label{sec:issues}
This section focuses on a few important questions about the validity and the scope of the TFL.  The most interesting of these may be the origin of the three-level Ansatz, which appears to be experimentally and theoretically valid.  Related to this is the nature of the limits themselves and how the sum rules and the Hamiltonian limits may be reconciled.  To this end, it is imperative to analytically explore the use of exotic Hamiltonians to determine their fundamental limits.  Monte Carlo simulations have discovered the optimum energy spectrum scaling, but not the optimum way to control eigenfunctions so that their transition moments are optimized.  A natural extension of the TFL is to determine the limits of composite operators, such as real($\beta$)/imag($\alpha$), which is a figure of merit (FOM) for an electro-optic device. Computation of the FOM of optical devices such as all-optical switches and modulators should lead to interesting new design rules. Each of these is discussed with the spirit of encouraging graduate students and researchers to tackle them.

\subsection{The \emph{limit} gap}\label{sec:issue1}
The existence of a Hamiltonian $H(r,p)$ satisfying the commutation relation $[r^{i},[r^{j},H]]=-(\hbar^2/m)\delta_{ij}$ leads directly to the TRK sum rules in Eqn \ref{TRKsumrule}.  The sets of spectra and transition moments $(E_{nm},x_{nm})$ computed from this Hamiltonian will satisfy the sum rules, and when they are optimized and used in the sum over states for the intrinsic $\beta$ or $\gamma$, they yield maximum intrinsic value of $\beta=0.7089$, while $-0.15\leq\gamma\leq 0.6$.  This holds for all potential models studied to date.

The selection of sets of spectra and transition moments $(E_{nm},x_{nm})$ at random but constrained by the TRK sum rules leads to a maximum intrinsic value of $\beta=1$ and $-0.25\leq\gamma\leq 1$.  This is true provided that the three-level Ansatz (TLA) is valid.  It is possible to violate these limits with pathological spectra which violate the TLA\cite{shafe13.01}, but these violations are nonphysical if the TLA is valid.  For the moment, we have no reason to question the validity of the TLA, as it holds in every model studied to date.

These two maxima for $\beta$ were displayed in Figure \ref{fig:betaLimitFig}, where the separation between them, the \emph{limit} gap, is evident.  An interesting theoretical question is whether the states and spectra of an exotic Hamiltonian will produce limits which lie in the limit gap.  Such a Hamiltonian will take the three dimensional form\cite{watki12.01}
\begin{eqnarray}\label{Hexotic}
H(\textbf{r},\textbf{p})&=&\frac{\textbf{p}^{2}}{2m}+d(\textbf{r})p_{x}p_{y}p_{z}+\sum_{i=1}^{3}\sum_{j=i+1}^{3}a_{ij}(\textbf{r})p_{i}p_{j}\nonumber \\
&+&\sum_{i=1}^{3}b_{i}(\textbf{r})p_{i}+V(\textbf{r})
\end{eqnarray}
where $\textbf{r}=[x,y,z]$, $\textbf{p}=[p_x,p_y,p_z]$, and $a,b,V$ are arbitrary tensor functions of the position operators with ranks indicated by their indices.  This is the most general Hamiltonian for which the TRK sum rules are valid. Note that the quantum mechanical version of this Hamiltonian must be made Hermitian before it is used to generate any physics.

\begin{figure}\centering
\includegraphics[width=3.2in]{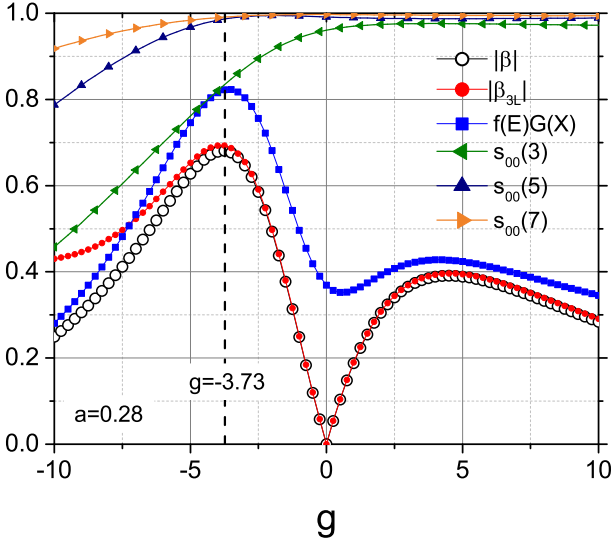}\\
\caption{First hyperpolarizability $\beta$, the three-level sum $\beta_{3L}$, and the three-level $\beta=f(E)G(X)$ for a one-dimensional $\delta$ atom, plotted against the strength $g$ of the $\delta$ potential.  Both $\beta$ and $\beta_{3L}$ converge near the maximum, whereas the three-level model overestimates the correct hyperpolarizability by over $20\%$  Also shown are the normalized partial sums $s_{00}\equiv (m/N\hbar^2)S_{00}(n)$ when truncated to level n.  The failure of the three-level model to accurately predict $\beta$ can be traced directly to the failure of $S_{00}(3)$ to converge with three levels.  The results show that the TLA is active when $\beta$ nears the Hamiltonian Limit, but that it does not apply to the sum rules.}\label{fig:1delta_beta3L_beta_fG_partialSums}
\end{figure}

In quantum graph models, it is found that the sum rules for a graph with a maximum near the Hamiltonian limit $\beta=0.7089$ has a $\beta$ function that differs from the  three-level model and only aligns with the three-level sum over states when $\beta$ is near that maximum.  Figure \ref{fig:1delta_beta3L_beta_fG_partialSums} illustrates this for the $\delta$ atom with potential $V(x)=(g/2L)\delta(x-a)$, and $a=0.28$.  The full three level model and the exact sum over states converge nearly everywhere above $g=-3.73$, where the maximum of $0.68$ is attained.  Below this value of g, the two expressions diverge slightly.  But the three level model is in error at the maximum by at least 30$\%$.  The failure of one of the sum rules $S_{00}$ used in the three-level model is shown in the Figure, where it is seen that this basic sum rule requires at least five levels to get near its exact value of unity.

The Figure illustrates both the operation of the TLA, and the fact that the TLA itself is not exact: The sum rules always require more states to converge to a few percent of their actual values, while the hyperpolarizabilities can get to within a few percent with only three ($\beta$) or four ($\gamma$) states. The results will be similar for any of the potential models discussed in Section \ref{sec:results}, none of which were exotic Hamiltonians.

The problem to be solved is then this:  Starting with an exotic Hamiltonian, solve for the spectra and eigenstates, construct the transition moments, compute $\beta$ using the sum over states, optimize the potential such that $\beta$ reaches its maximum, and see if it exceeds $0.7089$.  At the same time, study the convergence of the sum rules with three levels to see how close they get to their correct values.  This is an intellectual exercise, but the knowledge gained is significant.  The practical value may be limited, because if a newly synthesized molecule gets to the Hamiltonian Limit, we don't much care if it can get a little above that.  This statement may change someday if we learn how to routinely design molecules whose maximum intrinsic nonlinearities approach the limits.  That day may not be far off.

\subsection{Proving the three-level Ansatz}\label{sec:issue2}
The three-level Ansatz (TLA) states that only three levels are required when $\beta$ is near its maximum value.  The maximum to which this refers is the three-level model limit in Figure \ref{fig:betaLimitFig}. The actual limit predicted by the full sum rules is also this limit, provided that the TLA is actually valid.

A corollary to the above definition of the TLA is that only three levels are required when $\beta$ is near its local maximum value for a given system, even if the local maximum is far from the theoretical maximum.  All Hamiltonian systems studied to date, including the plethora of quantum graphs delineated in the prior section, satisfy the TLA when their optimum geometries are discovered, even though many fall short of the Hamiltonian Limit.  There is something fundamental about the way the TLA works.

The TLA has yet to be proven.  It may be exact, in which case the first three terms in the sum over states for $\beta$ in Eqn \ref{betaijk} converge to the exact maximum, and the rest sum to zero, or it may be approximate, \emph{nearly} converging to the limit but with some residual value from the remaining sum over states, all while the sum rules hold exactly.  A system with the maximum $\beta$ in the three-level model has $X=0.79$ and $E=0$, and is a system with wide level spacing between the first and second excited states.   But as we have seen, Hamiltonian systems often have $E\sim 0.4-0.5$, but also satisfy the TLA. The TLA was required to rescue the TFL from the plague of infinities in the many-state catastrophe\cite{shafe13.01}.

But where does it come from?  More to the point, which energy levels $E_{n0}$ and transition moments $x_{n0}$ cause $\beta-\beta_{3L}$ to converge to a small or zero value when this same set of spectra and moments satisfy the sum rules? This is a multi-dimensional minimization problem with an infinite set of constraint equations, the sum rules.  In fact, we can write
\begin{equation}\label{betaResidual}
\Delta\beta\equiv\beta -\beta_{3L}=3e^3\sum_{n=m}^{\infty}\sum_{m=3}^{\infty}\left[\beta_{nm}+\beta_{mn}-\delta_{nm}\beta_{nn}\right],
\end{equation}
with the definition
\begin{equation}\label{betanm}
\beta_{nm}\equiv\frac{x_{0n}\bar{x}_{nm}x_{n0}}{E_{n0}E_{m0}}
\end{equation}
and attempt to minimize $\Delta\beta$ subject to the sum rules in Eqn \ref{TRKsumrule} to derive a set of equations for the spectra and moments. Assuming the spectra actually follow a specific scaling law would generate a set of equations for the moments, but even if this is solvable, what Hamiltonian corresponds to these moments?  More challenging is the fact that the three most important states depend on the specific configuration of the molecule, and can change as this configuration is varied.  This could render the process intractable.

But the TLA is fundamental to the TFL so it must be proved.  Shafei and Kuzyk speculate that it is true but not provable\cite{shafe13.01}, much like the Godel theorem for axiomatic arithmatic systems.  Perhaps a solution will emerge in the coming years, but in the meantime, the TLA has not failed even once and is assumed by most researchers to be a universal principle.

\subsection{Design ideas from the TFL}\label{sec:issue3}
The Monte Carlo simulations described in Section \ref{sec:results} showed that the ideal spectrum for the states of a structure contributing the most to the nonlinearities scale quadratically or faster with mode number.  Molecules don't scale this way, in general, but conjugated chains can exhibit this behavior.  Even so, the ideal spectrum is necessary but not sufficient to generate a large response.  The eigenfunctions should generate large changes in dipole moment under excitation by a real or virtual photon, and at the same time, maintain large mode overlap.  This ensures that the numerators in the terms of the three-level expansion Eqn \ref{beta3L} are large.  A recent letter developed a new design paradigm, \emph{phase disruption}, for generating the optimum transition moments\cite{lytel15.02}.  The method was developed for N electron quantum graph systems, models of quasi-one dimensional structures such as networks of nanowires.  The ground state wavefunction along the main chain has a kink created by the presence of a short side chain or group, which diverts some electron flux from the main change and creates a disruption in the phase along the chain.  This, in turn creates a preferential charge distribution to one side of the prong.  The side group causes a large change in dipole moment upon excitation of an electron from the ground state, and for specific lengths of the chain, the mode overlap remains large.  Since the hyperpolarizability depends on the product of the change in dipole moment and mode overlap, phase disruption may lead to large enhancements in the nonlinear optical response.

Modulated conjugation\cite{perez07.01,perez09.01} systems were the first to leap the apparent limit shown in Figure \ref{fig:betaLimitFig}, but their nonlinearities remained well below the limit.  A cursory view of the work reveals that the eigenfunctions take on the general characteristics required for optimum transition moments, viz., shifting charge localization from the ground to first excited state (producing a large change in dipole moment) while maintaining some mode overlap and this enabling $\beta$ to increase.  If the spectra can be optimized, then this design method may lead to better molecules, as well.

To date, nearly all Hamiltonian and Monte Carlo studies of the TFL have been one-dimensional models.  Phase disruption and modulated conjugation are two promising methods for enhancing nonlinearities, but they may fail miserably in real molecules, simply because the spatial mode overlap required to enhance transition moments may be small.

\subsection{Application to device figures of merit}\label{sec:issue4}
The fundamental limit of a molecule is now established, and with it, limits on real materials may be computed by assembling aggregations, aligning if necessary to eliminate centrosymmetry, and computing local field factors.  But for devices, the relevant parameters are the macroscopic refractive index change with a low-frequency field (second-order electro-optic or magneto-optic effect), or with an intense light beam (third order).  The former has applications to electro-optic modulators, phase conjugators using diffracting gratings in spatial light modulators, beam steering, and filtering\cite{yariv77.01,yariv78.01}.  The latter has applications to light by light control for switching\cite{gibbs84.01}, frequency mixing\cite{maker65.02}, and harmonic generation\cite{bloem68.01}.  The figure of merit (FOM) of a device is in general a product or ratio of physically relevant parameters, such as the switching voltage or intensity, and the optical absorption.  These, in turn, are functions of the dispersive second- and third-order susceptibilities, calculated from the first and second hyperpolarizabilities.  A contributed paper in this volume shows how to compute the FOM for an electro-optic device. Similar computations for all-optical devices are clearly relevant and rather complex, and may be of interest to graduate students and researchers.  This is fertile ground for exploring materials science for device physics.

\subsection{What we know--and don't know--about the TFL today}\label{sec:issue5}
We end this section with a concise list of known and unknown features of the TFL:
\begin{enumerate}
\item
Truncating the sum over states \emph{and} sum rules to three levels produces the three-level model $\beta=\beta_{max}f(E)G(X)$, where $\beta_{max}$ is given in Eqn \ref{betaMax}.
\item
The three-level model violates the sum rules.  In particular, $S_{22}$ is explicitly violated.  Therefore, the moments computed using the truncated sum rules and substituted into the three-level sum over states to produce the three-level model are in error, no matter what system for which we specify an $x_{01}$ and energy level differences $E_{10}$ and $E_{20}$.
\item
If we truncate the sum over states to three levels but not the sum rules, we are simply stopping the sum over states after three levels using the moments and spectra we have in hand from some theory (e.g., a solved Hamiltonian), and we'll find that this truncated sum, called $\beta_{3L}$, is less than $\beta_{max}$.  We cannot prove this.  But we believe it is true.
\item
The full sum over states for $\beta$ approaches $\beta_{3L}$ when $\beta$ is nearing its maximum.  \emph{This is the statement of the three-level Ansatz}.
\item
Solving a Hamiltonian for the spectra and states (and calculating the moments) produces a number for $\beta$ and satisfies the sum rules.  But picking spectra and moments that satisfy the sum rules does not mean there is a Hamiltonian that could have generated them.
\item
If one truncates sum rules and the sum over states to N levels, then it is possible that $\beta_{max}$ computed this way will diverge as N goes to infinity.  But the states and spectra that cause this to happen violate the TLA.  So the TLA in effect rescues the three-level model.
\item
Finally, we know that to date, all Hamiltonian systems have $\beta\leq 0.7089$.  We also know that the Monte Carlo calculations generated some solutions with $0.7089\leq\beta\leq 1$, but we don't know that the spectra and moments that cause $\beta$ to get into this range are physical.
\end{enumerate}

It would be equally informative to list what we don't know about the TFL today:
\begin{enumerate}
\item
We can't explain the limit gap.  The Monte Carlo calculations reviewed in this paper used exact sum rules to constrain otherwise random sets $(E_{n0},x_{nm})$ of spectra and moments. Constrained by the TLA, these models hit the Three-level Model Limit in Table \ref{tab:modelLimits}. Perhaps the cardinality of all the real number pairs selected this way is larger than that of all pairs that could be generated by a physical Hamiltonian, which appear to be bounded by the Hamiltonian Limits.  Spectral theory of Hamiltonian operators, often studied in quantum chaos, is probably required to make any inroads into understanding the limit gap.  This remains to be investigated.
\item
We have no idea why the numerical value of the Hamiltonian limit is what it is.  What is special about $0.7089$?
\item
We can't yet specify a design (potential, symmetry, dimensionality) that will hit the limits.
\item
We have yet to study potential optimization models of three-dimensional, interacting multi-electron systems that actually resemble a molecule.
\item
We don't know which limits apply to interacting multielectron, multidimensional systems, as the three-level model limit is an approximation whose value as an upper bound to an actual limit for such systems remains unproven.
\item
We haven't proved the three-level Ansatz.
\end{enumerate}

The TFL was invented nearly sixteen years ago and has led to discoveries of better molecules, numerous analytical tests, and a deep understanding of the origins of nonlinear optical effects in molecular-scale structures. The field should continue to spawn significant new studies and many thesis topics.

\section{Conclusions}\label{sec:end}
The invention of the Erbium-doped fiber amplifier, the advent of high speed modulation of laser light, and the realization of dense wavelength division multiplexers led to the construction of an optical network hardware layer capable of supporting the explosion of the internet and the world-wide web in the mid 1990s, providing the bandwidth to support TCP/IP on a global scale.  Google, Facebook, and other search and social media enterprises now dominate the business landscape and have displaced IBM and AT\&T as the darlings of the stock market.  Though computing itself is still dominated by electronic devices, terrestrial and sub-oceanic communication is optical.  We all depend upon the ubiquity of optical devices in our professional and personal lives.  It is exciting for a scientist to think that a few advances in optical materials science could have such a large impact on human life.

Nonlinear optics has yet to achieve this ubiquity, but may do so once more responsive materials are created.  The theory of fundamental limits (TFL) has revealed that present molecular designs may be much improved to enhance their optical nonlinearities.  Explorations described in the previous section suggest that the energy spectra must be optimized to scale quadratically or faster with mode number, at least for the low lying levels that contribute the most to $\beta$ and $\gamma$.  Topological and geometrical optimizations of spectrally acceptable systems can then lead to greater enhancements, but if the spectra don't scale well, then neither will the nonlinearities, no matter how their symmetries and shapes are adjusted.

This review has identified a number of fundamentally important and interesting theoretical problems that students and researchers can address in the coming years: (1) Discover the fundamental principles behind the three-level Ansatz and offer a proof, (2) Solve a model with an exotic Hamiltonian and optimize the hyperpolarizabilities in an attempt to bridge the limit gap, (3) Solve a 2D and 3D model and optimize the hyperpolarizabilities to discover the effect of dimensionality, and (4) Compute the fundamental limits of figures of merit of optical devices to seek new operating conditions that maximize performance by trading off response and loss.  As an interested party, I wish ambitious students the best and expect them to drive the field forward with new discoveries and surprises.


\begin{thebibliography}{100}
\newcommand{\enquote}[1]{``#1''}

\bibitem{garmi13.01}
E.~Garmire, \enquote{Nonlinear optics in daily life,} Optics express
  \textbf{21}, 30532--30544 (2013).

\bibitem{baugh78.06}
R.~H. Baughman and R.~R. Chance, \enquote{{Fully Conjugated Polymer Crystalls:
  Solid-State Synthesis and Properties of the Polydiacetylenes},} Annals of the
  New York Academy of Sciences \textbf{313}, 705--725 (1978).

\bibitem{lipsc81.01}
G.~F. Lipscomb, A.~F. Garito, and R.~S. Narang, \enquote{{An exceptionally
  large linear electro-optic effect in the organic-solid MNA},} J. Chem Phys.
  \textbf{75}, 1509--1516 (1981).

\bibitem{hugga87.01}
P.~G. Huggard, W.~Blau, and D.~Schweitzer, \enquote{{Large Third-Order Optical
  Nonlinearity of the Organic Metal x-[Bis(Ethylenedithio) Tetrathiofulvalene]
  Triiodide},} Appl. Phys.Lett. \textbf{51}, 2183--5 (1987).

\bibitem{cheml80.01}
D.~S. Chemla, \enquote{{Non-Linear Optical Properties of Condensed Matter},}
  Rep. Prog. Phys. \textbf{43}, 1191--1262 (1980).

\bibitem{giord65.01}
J.~A. Giordmaine, \enquote{{Nonlinear Optical Properties of Liquids},} Phys.
  Rev. \textbf{138}, A 1599-- A 1606 (1965).

\bibitem{giord67.01}
J.~A. Giordmaine, \enquote{{Nonlinear Optical Properties of Liquids},} J. Chem.
  Phys. \textbf{64}, A1599--A1606 (1967).

\bibitem{ho79.01}
P.~P. Ho and R.~R. Alfano, \enquote{{Optical Kerr Effect in Liquids},} Phys.
  Rev. A \textbf{20}, 2170--87 (1979).

\bibitem{chen88.01}
C.~H. Chen and M.~P. McCann, \enquote{{Measurements of Two-photon Absorption
  Cross Sections for Liquid Benzene and Methyl Benzene},} J. Chem. Phys.
  \textbf{88}, 4671--7 (1988).

\bibitem{barni83.01}
M.~I. Barnik, L.~M. Blinov, A.~M. Dorozhkin, and N.~M. Shtykov,
  \enquote{{Optical Second Harmonic Generation in Various Liquid Crystalline
  Phases},} Mol. Cryst. Liq. Cryst. \textbf{98}, 1--12 (1983).

\bibitem{shelt82.01}
D.~P. Shelton and A.~D. Bucking, \enquote{{Optical Second-Harmonic Generation
  in Gases with a Low-Power Laser},} Phys. Rev. A \textbf{26}, 2787--2798
  (1982).

\bibitem{kaatz98.01}
P.~Kaatz, E.~A. Donley, and D.~P. Shelton, \enquote{{A comparison of molecular
  hyperpolarizabilities from gas and liquid phase measurements},} J. Chem.
  Phys. \textbf{108}, 849--856 (1998).

\bibitem{guble99.01}
U.~Gubler, C.~Bosshard, P.~Gunter, M.~Y. Balakina, J.~Cornil, J.~L. Bredas,
  R.~E. Martin, and F.~Diederich, \enquote{{Scaling law for
  second-orderhyperpolarizability in poly(triacetylene) molecular wires},} Opt.
  Lett. \textbf{24}, 1599 (1999).

\bibitem{foste08.01}
M.~Foster, A.~Turner, M.~Lipson, and A.~Gaeta, \enquote{Nonlinear optics in
  photonic nanowires,} Opt. Exp., \textbf{16}, 1300--1320 (2008).

\bibitem{tian09.01}
B.~Tian, P.~Xie, T.~Kempa, D.~Bell, and C.~Lieber, \enquote{{Single-crystalline
  kinked semiconductor nanowire superstructures},} Nat. Nanotech. \textbf{4},
  824--829 (2009).

\bibitem{cheml85.01}
D.~Chemla and D.~Miller, \enquote{Room-temperature excitonic nonlinear-optical
  effects in semiconductor quantum-well structures,} J. Opt. Soc. Am. B
  \textbf{2}, 1155--1173 (1985).

\bibitem{miller1984band}
D.~Miller, D.~Chemla, T.~Damen, A.~Gossard, W.~Wiegmann, T.~Wood, and
  C.~Burrus, \enquote{{Band-edge electroabsorption in quantum well structures:
  The quantum-confined Stark effect},} Physical Review Letters \textbf{53},
  2173--2176 (1984).

\bibitem{rink89.01}
S.~Schmitt-Rink, D.~S. Chemla, and D.~A.~B. Miller, \enquote{{Linear and
  Nonlinear Optical Properties of Semiconductor Quantum Wellls},} Adv. Phys.
  \textbf{38}, 89--188 (1989).

\bibitem{schmi87.01}
S.~Schmitt-Rink, D.~Miller, and D.~Chemla, \enquote{Theory of the linear and
  nonlinear optical properties of semiconductor microcrystallites,} Phys. Rev.
  B \textbf{35}, 8113 (1987).

\bibitem{kuzyk13.02}
M.~G. Kuzyk, K.~D. Singer, and G.~Stegeman, \enquote{Theory of molecular
  nonlinear optics,} Advances in Optics and Photonics \textbf{5}, 4--82 (2013).

\bibitem{kuzyk14.02}
M.~G. Kuzyk, \emph{Lecture Notes in Nonlinear Optics: A student's perspective}
  (NLOsource, 2014).

\bibitem{kuzyk00.01}
M.~G. Kuzyk, \enquote{{Physical Limits on Electronic Nonlinear Molecular
  Susceptibilities},} Phys. Rev. Lett. \textbf{85}, 1218 (2000).

\bibitem{mille81.01}
A.~Miller, D.~A. Miller, and S.~D. Smith, \enquote{Dynamic non-linear optical
  processes in semiconductors,} Advances in Physics \textbf{30}, 697--800
  (1981).

\bibitem{kuzyk94.01}
M.~G. Kuzyk and C.~Poga, \emph{Molecular Nonlinear Optics: Materials, Physics,
  and Devices} (Academic Press, San Diego, 1994), chap. Quadratic
  Electro-Optics of Huest-Host Polymers, pp. 299--337, Quantum Electronics -
  Priciples and Applications.

\bibitem{kuzyk06.06}
M.~G. Kuzyk, \emph{Polymer Fiber Optics: materials, physics, and applications},
  vol. 117 of \emph{Optical science and engineering} (CRC Press, Boca Raton,
  2006).

\bibitem{kuzyk00.02}
M.~G. Kuzyk, \enquote{{Fundamental limits on third-order molecular
  susceptibilities},} Opt. Lett. \textbf{25}, 1183 (2000).

\bibitem{orr71.01}
B.~J. Orr and J.~F. Ward, \enquote{{Perturbation Theory of the Non-Linear
  Optical Polarization of an Isolated System},} Molec. Phys. \textbf{20},
  513--526 (1971).

\bibitem{lytel13.01}
R.~Lytel, S.~Shafei, J.~H. Smith, and M.~G. Kuzyk, \enquote{Influence of
  geometry and topology of quantum graphs on their nonlinear optical
  properties,} Phys. Rev. A \textbf{87}, 043824 (2013).

\bibitem{schif68.01}
L.~I. Schiff, \emph{Quantum Mechanics}, International Pure and Applied Physics
  Series (McGraw-Hill, New York, 1968), 3rd ed.

\bibitem{kuzyk03.02}
M.~G. Kuzyk, \enquote{{Erratum: Physical Limits on Electronic Nonlinear
  Molecular Susceptibilities},} Phys. Rev. Lett. \textbf{90}, 039902 (2003).

\bibitem{thom25.01}
W.~Thomas, \enquote{\"{U}ber die zahl der dispersionselektronen, die einem
  station{\\"a}ren zustande zugeordnet sind.(vorlaufige mitteilung),}
  Naturwissenschaften \textbf{13}, 627--627 (1925).

\bibitem{reich25.01}
F.~Reiche and u.~W. Thomas, \enquote{\"{U}ber die zahl der
  dispersionselektronen, die einem station{\"a}ren zustand zugeordnet sind,} Z.
  Physik \textbf{34}, 879 (1925).

\bibitem{kuhn25.01}
W.~Kuhn, \enquote{\"{U}ber die gesamtstarke der von einem zustande ausgehenden
  absorptionslinien,} Zeitschrift fur Physik A: Hadrons and Nuclei \textbf{33},
  408--412 (1925).

\bibitem{Bethe77.01}
H.~Bethe and E.~Salpeter, \emph{Quantum mechanics of one-and two-electron
  atoms} (Plenum Publishing Corporation, 1977).

\bibitem{ferna02.01}
F.~Fern{\'a}ndez, \enquote{The thomas reiche kuhn sum rule for the rigid
  rotator,} Int. J. Math. Ed. Sc. Tech. \textbf{33}, 636--640 (2002).

\bibitem{watki12.01}
D.~Watkins and M.~Kuzyk, \enquote{Universal properties of the optimized
  off-resonant intrinsic second hyperpolarizability,} J. Opt. Soc. Am. B
  \textbf{29}, 1661--1671 (2012).

\bibitem{kuzyk13.01}
M.~G. Kuzyk, J.~Perez-Moreno, and S.~Shafei, \enquote{Sum rules and scaling in
  nonlinear optics,} Phys. Rep \textbf{529}, 297--398 (2013).

\bibitem{champ05.01}
B.~Champagne and B.~Kirtman, \enquote{{Comment on "Physical Limits on
  Electronic Nonlinear Molecular Susceptibilities"},} Phys. Rev. Lett.
  \textbf{95}, 109401 (2005).

\bibitem{kuzyk05.01}
M.~G. Kuzyk, \enquote{{Reply to comment on "Physical Limits on Electronic
  Nonlinear Molecular Susceptibilities"},} Phys. Rev. Lett. \textbf{95}, 109402
  (2005).

\bibitem{kuzyk14.01}
M.~G. Kuzyk, \enquote{A heuristic approach for treating pathologies of
  truncated sum rules in limit theory of nonlinear susceptibilities,} arXiv
  preprint arXiv:1402.3827  (2014).

\bibitem{shafe13.01}
S.~Shafei and M.~G. Kuzyk, \enquote{Paradox of the many-state catastrophe of
  fundamental limits and the three-state conjecture,} Phys. Rev. A \textbf{88},
  023863 (2013).

\bibitem{singe98.01}
K.~Singer, S.~Hubbard, A.~Schober, L.~Hayden, and K.~Johnson,
  \enquote{Characterization techniques and tabulations for organic nonlinear
  optical materials, chapter 6; kuzyk, mg, dirk, cw 1998,} .

\bibitem{tripa04.01}
K.~Tripathy, J.~Perez-Moreno, M.~G. Kuzyk, B.~J. Coe, K.~Clays, and A.~M.
  Kelley, \enquote{Why hyperpolarizabilities fall short of the fundamental
  quantum limits,} J. Chem. Phys. \textbf{121}, 7932 (2004).

\bibitem{shafe11.01}
S.~Shafei and M.~G. Kuzyk, \enquote{Critical role of the energy spectrum in
  determining the nonlinear-optical response of a quantum system,} J. Opt. Soc.
  Am. B \textbf{28}, 882--891 (2011).

\bibitem{kuzyk06.03}
M.~G. Kuzyk, \enquote{{Fundamental limits of all nonlinear-optical phenomena
  that are representable by a second-order susceptibility},} J. Chem Phys.
  \textbf{125}, 154108 (2006).

\bibitem{kuzyk05.02}
M.~G. Kuzyk, \enquote{{Compact sum-over-states expression without dipolar terms
  for calculating nonlinear susceptibilities},} Phys. Rev. A \textbf{72},
  053819 (2005).

\bibitem{leung86.01}
P.~T. Leung and M.~L. Rustgi, \enquote{Relativistic corrections to bethe sum
  rule,} Phys. Rev. A \textbf{33} (1986).

\bibitem{cohen98.01}
S.~Cohen and P.~Leung, \enquote{General formulation of the semirelativistic
  approach to atomic sum rules,} Physical Review A \textbf{57}, 4994 (1998).

\bibitem{sinky06.01}
H.~Sinky and P.~Leung, \enquote{Relativistic corrections to a generalized sum
  rule,} Physical Review A \textbf{74}, 034703 (2006).

\bibitem{dawson15.01}
N.~J. Dawson, \enquote{Lowest-order relativistic corrections to the fundamental
  limits of nonlinear-optical coefficients,} Physical Review A \textbf{91},
  013832 (2015).

\bibitem{jacki67.01}
R.~Jackiw, \enquote{Quantum-mechanical sum rules,} Physical Review
  \textbf{157}, 1220 (1967).

\bibitem{wang99.01}
S.~Wang, \enquote{Generalization of the thomas-reiche-kuhn and the bethe sum
  rules,} Phys. Rev. A \textbf{60}, 262--266 (1999).

\bibitem{bassa91.01}
F.~Bassani and S.~Scandolo, \enquote{Dispersion relations and sum rules in
  nonlinear optics,} Physical Review B \textbf{44}, 8446 (1991).

\bibitem{scand92.01}
S.~Scandolo and F.~Bassani, \enquote{Nonlinear sum rules: The three-level and
  the anharmonic-oscillator models,} Physical Review B \textbf{45}, 13257
  (1992).

\bibitem{mavro93.01}
H.~Mavromatis, \enquote{Sum rules, as a tool for obtaining mathematical
  series,} Int. J. Math. Educ. Sci. Tech \textbf{26}, 267--313 (1993).

\bibitem{clays01.01}
K.~Clays, \enquote{{Theoretical upper limits and experimental overestimates for
  molecular hyperpolarizabilities: a symbiosis},} Opt. Lett. \textbf{26},
  1699--1701 (2001).

\bibitem{clays03.01}
K.~Clays and B.~J. Coe, \enquote{{Design strategies versus limiting theory for
  engineering large second-order nonlinear optical polarizabilities in charged
  organic molecules},} Chem. Mater. \textbf{15}, 642--648 (2003).

\bibitem{kuzyk03.03}
M.~G. Kuzyk, \enquote{{Fundamental Limits on Two-Photon Absorption
  Cross-Sections},} J. Chem Phys. \textbf{119}, 8327--8334 (2003).

\bibitem{slepk04.01}
A.~D. Slepkov, F.~A. Hegmann, S.~Eisler, E.~Elliot, and R.~R. Tykwinski,
  \enquote{{The surprising nonlinear optical properties of conjugated polyyne
  oligomers},} J. Chem. Phys. \textbf{120}, 6807--6810 (2004).

\bibitem{champ06.01}
B.~Champagne and B.~Kirtman, \enquote{{Evaluation of alternative
  sum-over-states expressions for the first hyperpolarizability of push-pull
  pi-conjugated systems},} J . Chem. Phys. \textbf{125}, 024101 (2006).

\bibitem{kuzyk08.01}
M.~C. Kuzyk and M.~G. Kuzyk, \enquote{{Monte Carlo Studies of the Fundamental
  Limits of the Intrinsic Hyperpolarizability},} J. Opt. Soc. Am. B.
  \textbf{25}, 103--110 (2008).

\bibitem{shafe10.01}
S.~Shafei, M.~C. Kuzyk, and M.~G. Kuzyk, \enquote{Monte carlo studies of the
  intrinsic second hyperpolarizability,} J. Opt. Soc Am. B \textbf{27},
  1849--1856 (2010).

\bibitem{zhou06.01}
J.~Zhou, M.~G. Kuzyk, and D.~S. Watkins, \enquote{{Pushing the
  hyperpolarizability to the limit},} Opt. Lett. \textbf{31}, 2891 (2006).

\bibitem{perez07.01}
J.~P\'{e}rez-Moreno, Y.~Zhao, K.~Clays, and M.~G. Kuzyk, \enquote{{Modulated
  conjugation as a means for attaining a record high intrinsic
  hyperpolarizability},} Opt. Lett. \textbf{32}, 59--61 (2007).

\bibitem{perez07.02}
J.~P\'{e}rez-Moreno, I.~Asselberghs, Y.~Zhao, K.~Song, H.~Nakanishi, S.~Okada,
  K.~Nogi, O.-K. Kim, J.~Je, J.~Matrai, M.~De~Mayer, and M.~G. Kuzyk,
  \enquote{{Combined molecular and supramolecular bottom-up nano-engineering
  for enhanced nonlinear optical response: Experiments, modelling and
  approaching the fundamental limit},} J. Chem. Phys. \textbf{126}, 074705
  (2007).

\bibitem{perez09.01}
J.~P\'{e}rez-Moreno, Y.~Zhao, K.~Clays, M.~G. Kuzyk, Y.~Shen, L.~Qiu, J.~Hao,
  and K.~Guo, \enquote{Modulated conjugation as a means of improving the
  intrinsic hyperpolarizability,} J. Am. Chem. Soc. \textbf{131}, 5084--5093
  (2009).

\bibitem{may05.01}
J.~C. May, J.~H. Lim, I.~Biaggio, N.~N.~P. Moonen, T.~Michinobu, and
  F.~Diederich, \enquote{{Highly efficient third-order optical nonlinearities
  in donor-substituted cyanoethynylethene molecules},} Opt. Lett. \textbf{30},
  3057--3059 (2005).

\bibitem{may07.01}
J.~C. May, I.~Biaggio, F.~Bures, and F.~Diederich, \enquote{{Extended
  conjugation and donor-acceptor substitution to improve the third-order
  optical nonlinearity of small molecules},} App. Phys. Lett. \textbf{90},
  251106 (2007).

\bibitem{kang05.01}
H.~Kang, A.~Facchetti, P.~Zhu, H.~Jiang, Y.~Yang, E.~Cariati, S.~Righetto,
  R.~Ugo, C.~Zuccaccia, A.~Macchioni, C.~L. Stern, Z.~Liu, S.~T. Ho, and T.~J.
  Marks, \enquote{{Exceptional Molecular Hyperpolarizabilities in Twisted
  $\pi$-Electron System Chromophores},} Angew. Chem. Int. Ed. \textbf{44},
  7922--7925 (2005).

\bibitem{kang07.01}
H.~Kang, A.~Facchetti, H.~Jiang, E.~Cariati, S.~Righetto, R.~Ugo, C.~Zuccaccia,
  A.~Macchioni, C.~L. Stern, Z.~F. Liu, S.~T. Ho, E.~C. Brown, M.~A. Ratner,
  and T.~J. Marks, \enquote{{Ultralarge hyperpolarizability twisted
  $\pi$-electron system electro-optic chromophores: Synthesis, solid-state and
  solution-phase structural characteristics, electronic structures, linear and
  nonlinear optical properties, and computational studies},} J. Am. Chem. Soc.
  \textbf{129}, 3267--3286 (2007).

\bibitem{he11.01}
G.~S. He, J.~Zhu, A.~Baev, M.~Samoc{\`I}, D.~L. Frattarelli, N.~Watanabe,
  A.~Facchetti, H.~{\~A}{\ldots}gren, T.~J. Marks, and P.~N. Prasad,
  \enquote{Twisted $\pi$-system chromophores for all-optical switching,} J. Am.
  Chem. Soc. \textbf{133}, 6675--6680 (2011).

\bibitem{zhou08.01}
J.~Zhou and M.~G. Kuzyk, \enquote{{Intrinsic Hyperpolarizabilities as a Figure
  of Merit for Electro-optic Molecules},} J. Phys. Chem. C. \textbf{112},
  7978--7982 (2008).

\bibitem{perez11.02}
J.~P\'{e}rez-Moreno, H.~S.-T., M.~G. Kuyzk, Z.~Zhou, S.~K. Ramini, and
  K.~Clays, \enquote{Experimental verification of a self-consistent theory of
  the first-, second-, and third-order (non)linear optical response,} Phys.
  Rev. A \textbf{84}, 033837 (2011).

\bibitem{jiang12.01}
N.~Jiang, G.~Zuber, S.~Keinan, A.~Nayak, W.~Yang, M.~Therien, and D.~Beratan,
  \enquote{Design of coupled porphyrin chromophores with unusually large
  hyperpolarizabilities,} J. Phys. Chem. C  (2012).

\bibitem{cleuv14.01}
S.~van Cleuvenbergen, I.~Asselberghs, W.~Vanormelingen, T.~Verbiest, E.~Franz,
  K.~Clays, M.~G. Kuzyk, and G.~Koeckelberghs, \enquote{Record-high
  hyperpolarizabilities in conjugated polymers,} J Mat. Chem. C 2 \textbf{2},
  4533 (2014).

\bibitem{al2016donor}
A.~Al-Yasari, N.~Van~Steerteghem, H.~El~Moll, K.~Clays, and J.~Fielden,
  \enquote{Donor-acceptor organo-imido polyoxometalates: High transparency,
  high activity redox-active nlo chromophores,} Dalton Transactions  (2016).

\bibitem{coe2016helquat}
B.~J. Coe, D.~Rusanova, V.~D. Joshi, S.~S{\'a}nchez, J.~Vavra, D.~Khobragade,
  L.~Severa, I.~Cisarova, D.~Saman, R.~Pohl \emph{et~al.}, \enquote{Helquat
  dyes: Helicene-like push-pull systems with large second-order nonlinear
  optical responses,} The Journal of organic chemistry  (2016).

\bibitem{perez16.01}
J.~Perez-Moreno, S.~Shafei, and M.~G. Kuzyk, \enquote{Using universal scaling
  laws to identify the best molecular design paradigms for second-order
  nonlinear optics,} Physics arXiv \textbf{1604.03846} (2016).

\bibitem{perez16.02}
S.~K. M.~G. Perez-Moreno, J.;~Shafei, \enquote{Applying universal scaling laws
  to identify the best molecular design paradigms for third-order nonlinear
  optics,} arXiv:1604.03779  (2016).

\bibitem{dalga55.01}
A.~Dalgarno and J.~T. Lewis, \enquote{The exact calculation of long-range
  forces between atoms by perturbation theory,} Proc. R. Soc. London Ser. A
  \textbf{233}, 70--74 (1955).

\bibitem{kuzyk06.02}
M.~G. Kuzyk and D.~S. Watkins, \enquote{{The Effects of Geometry on the
  Hyperpolarizability},} J. Chem Phys. \textbf{124}, 244104 (2006).

\bibitem{zhou07.02}
J.~Zhou, U.~B. Szafruga, D.~S. Watkins, and M.~G. Kuzyk, \enquote{{Optimizing
  potential energy functions for maximal intrinsic hyperpolarizability},} Phys.
  Rev. A \textbf{76}, 053831 (2007).

\bibitem{watki11.01}
D.~S. Watkins and M.~G. Kuzyk, \enquote{The effect of electron interactions on
  the universal properties of systems with optimized off-resonant intrinsic
  hyperpolarizability,} J. Chem. Phys. \textbf{134}, 094109 (2011).

\bibitem{lytel15.02}
R.~{Lytel}, S.~{Mossman}, and M.~{Kuzyk}, \enquote{Phase disruption as a new
  design paradigm for optimizing the nonlinear-optical response,} Optics
  Letters \textbf{40}, 4735--4738 (2015).

\bibitem{watki09.01}
D.~S. Watkins and M.~G. Kuzyk, \enquote{Optimizing the hyperpolarizability
  tensor using external electromagnetic fields and nuclear placement,} J. Chem.
  Phys. \textbf{131}, 064110 (2009).

\bibitem{ather12.01}
T.~Atherton, J.~Lesnefsky, G.~Wiggers, and R.~Petschek, \enquote{Maximizing the
  hyperpolarizability poorly determines the potential,} J. Opt. Soc. Am. B
  \textbf{29}, 513--520 (2012).

\bibitem{burke13.01}
C.~J. Burke, T.~J. Atherton, J.~Lesnefsky, and R.~G. Petschek,
  \enquote{Optimizing the second hyperpolarizability with minimally
  parametrized potentials,} J. Opt. Soc. Am. B \textbf{30}, 1438--1445 (2013).

\bibitem{burke16.01}
C.~J. Burke, J.~Lesnefsky, R.~G. Petschek, and T.~J. Atherton,
  \enquote{Maximizing the hyperpolarizability of 1d potentials with multiple
  electrons,} arXiv preprint arXiv:1602.05246  (2016).

\bibitem{pauli36.01}
L.~Pauling, \enquote{The diamagnetic anisotropy of aromatic molecules,} J.
  Chem. Phys. \textbf{4}, 673 (1936).

\bibitem{kuhn48.01}
H.~Kuhn, \enquote{{Free Electron Model for Absorption Spectra of Organic
  Dyes},} J. Chem. Phys. \textbf{16}, 840--841 (1948).

\bibitem{ruede53.01}
K.~Ruedenberg and C.~W. Scherr, \enquote{Free-electron network model for
  conjugated systems. i. theory,} The Journal of Chemical Physics \textbf{21},
  1565--1581 (1953).

\bibitem{scher53.01}
C.~W. Scherr, \enquote{Free-electron network model for conjugated systems. ii.
  numerical calculations,} The Journal of Chemical Physics \textbf{21},
  1582--1596 (1953).

\bibitem{platt53.01}
J.~R. Platt, \enquote{Free-electron network model for conjugated systems. iii.
  a demonstration model showing bond order and``free valence''in conjugated
  hydrocarbons,} The Journal of Chemical Physics \textbf{21}, 1597--1600
  (1953).

\bibitem{kowal90.01}
D.~Kowal, U.~Sivan, O.~Entin-Wohlman, and Y.~Imry, \enquote{Transmission
  through multiply-connected wire systems,} Physical Review B \textbf{42}, 9009
  (1990).

\bibitem{flesi87.01}
C.~Flesia, R.~Johnston, and H.~Kunz, \enquote{Strong localization of classical
  waves: a numerical study,} EPL (Europhysics Letters) \textbf{3}, 497 (1987).

\bibitem{ivche98.01}
E.~Ivchenko and A.~Kiselev, \enquote{Electron g factor in quantum wires and
  quantum dots,} Journal of Experimental and Theoretical Physics Letters
  \textbf{67}, 43--47 (1998).

\bibitem{sanch98.01}
J.~S{\'a}nchez-Gil, V.~Freilikher, I.~Yurkevich, and A.~Maradudin,
  \enquote{Coexistence of ballistic transport, diffusion, and localization in
  surface disordered waveguides,} Physical review letters \textbf{80}, 948
  (1998).

\bibitem{avish92.01}
Y.~Avishai and J.~Luck, \enquote{Quantum percolation and ballistic conductance
  on a lattice of wires,} Physical Review B \textbf{45}, 1074 (1992).

\bibitem{amovi04.01}
C.~Amovilli, F.~E. Leys, and N.~H. March, \enquote{Electronic energy spectrum
  of two-dimensional solids and a chain of c atoms from a quantum network
  model,} Journal of mathematical chemistry \textbf{36}, 93--112 (2004).

\bibitem{leys04.01}
F.~E. Leys, C.~Amovilli, and N.~H. March, \enquote{Topology, connectivity, and
  electronic structure of c and b cages and the corresponding nanotubes,}
  Journal of chemical information and computer sciences \textbf{44}, 122--135
  (2004).

\bibitem{kuchm07.01}
P.~Kuchment and O.~Post, \enquote{On the spectra of carbon nano-structures,}
  Communications in Mathematical Physics \textbf{275}, 805--826 (2007).

\bibitem{kotto97.01}
T.~Kottos and U.~Smilansky, \enquote{Quantum chaos on graphs,} Phys. Rev. Lett.
  \textbf{79}, 4794--4797 (1997).

\bibitem{kotto99.02}
T.~Kottos and U.~Smilansky, \enquote{Periodic orbit theory and spectral
  statistics for quantum graphs,} Ann. Phys. \textbf{274}, 76--124 (1999).

\bibitem{kotto00.01}
T.~Kottos and U.~Smilansky, \enquote{Chaotic scattering on graphs,} Physical
  review letters \textbf{85}, 968 (2000).

\bibitem{blume01.04}
R.~Bl{\"u}mel, Y.~Dabaghian, and R.~Jensen, \enquote{One-dimensional quantum
  chaos: explicitly solvable cases,} JETP Lett \textbf{74}, 258 (2001).

\bibitem{hul04.01}
O.~Hul, S.~Bauch, P.~Pakonski, N.~Savytskyy, K.~Zyczkowski, and L.~Sirko,
  \enquote{Experimental simulation of quantum graphs by microwave networks,}
  Physical Review E \textbf{69}, 056205 (2004).

\bibitem{shafe11.02}
S.~Shafei and M.~G. Kuzyk, \enquote{The effect of extreme confinement on the
  nonlinear-optical response of quantum wires,} J. Nonl. Opt. Phys. Mat.
  \textbf{20}, 427--441 (2011).

\bibitem{shafe12.01}
S.~Shafei, R.~Lytel, and M.~G. Kuzyk, \enquote{Geometry-controlled nonlinear
  optical response of quantum graphs,} J. Opt. Soc. Am. B \textbf{29},
  3419--3428 (2012).

\bibitem{lytel12.01}
R.~Lytel, S.~Shafei, and M.~G. Kuzyk, \enquote{Nonlinear optics of quantum
  graphs,} in \enquote{Proc. SPIE 8474,}  (2012), pp. 84740O--84740O--9.

\bibitem{lytel13.03}
R.~Lytel, S.~Shafei, and M.~G. Kuzyk, \enquote{Topological optimization of
  nonlinear optical quantum wire networks,} in \enquote{SPIE Organic Photonics+
  Electronics,}  (International Society for Optics and Photonics, 2013), pp.
  882702--882702.

\bibitem{lytel13.04}
R.~Lytel and M.~G. Kuzyk, \enquote{Dressed quantum graphs with optical
  nonlinearities approaching the fundamental limit,} Journal of Nonlinear
  Optical Physics \& Materials \textbf{22} (2013).

\bibitem{lytel14.01}
R.~Lytel, S.~Shafei, and M.~G. Kuzyk, \enquote{Optimum topology of
  quasi-one-dimensional nonlinear optical quantum systems,} Journal of
  Nonlinear Optical Physics \& Materials \textbf{23}, 1450025 (2014).

\bibitem{lytel15.01}
R.~Lytel, S.~M. {Mossman}, and M.~G. {Kuzyk}, \enquote{Optimization of
  eigenstates and spectra for quasi-linear nonlinear optical systems,} Journal
  of Nonlinear Optical Physics \& Materials \textbf{24}, 1550018 (2015).

\bibitem{blume06.01}
R.~Bl{\"u}mel, \enquote{Analytical solution of the compressed, one-dimensional
  delta atom via quadratures and exact, absolutely convergent periodic-orbit
  expansions,} Journal of Physics A: Mathematical and General \textbf{39}, 8257
  (2006).

\bibitem{yariv77.01}
A.~Yariv and D.~Pepper, \enquote{Amplified reflection, phase conjugation, and
  oscillation in degenerate four-wave mixing,} Optics letters \textbf{1},
  16--18 (1977).

\bibitem{yariv78.01}
A.~Yariv, \enquote{Phase conjugate optics and real-time holography,} IEEE J.
  Quant. Elec. \textbf{14}, 650--660 (1978).

\bibitem{gibbs84.01}
H.~Gibbs, \enquote{Physics of optical bistability,} J. Opt. Soc. Am. A, vol. 1,
  page 1281 \textbf{1}, 1281 (1984).

\bibitem{maker65.02}
P.~D. Maker and R.~W. Terhune, \enquote{{Study of Optical Effects Due to an
  Induced Polarization Third-Order in the Electric Field Strength},} Phys. Rev.
  \textbf{137}, 801--818 (1965).

\bibitem{bloem68.01}
N.~Bloembergen, R.~K. Chang, S.~S. Jha, and C.~H. Lee, \enquote{{Optical
  Second-Harmonic Generation in Reflection from Media with Inversion
  Symmetry},} Phys. Rev. \textbf{174}, 813--22 (1968).

\end{thebibliography}

\end{document}